%% file: paper.tex
\documentclass[conference]{IEEEtran}
\usepackage{fancyhdr}
\usepackage{cite}
\usepackage{amsmath,amssymb,amsfonts}
\usepackage{algorithmic}
\usepackage{graphicx}
\usepackage{textcomp}
\usepackage{xcolor}
\usepackage{booktabs}
\usepackage{multirow}
\usepackage{tikz}
\usepackage{subcaption}
\usepackage[hidelinks]{hyperref}
\usepackage{orcidlink}
\usepackage{pdfpages}
\def\BibTeX{{\rm B\kern-.05em{\sc i\kern-.025em b}\kern-.08em
    T\kern-.1667em\lower.7ex\hbox{E}\kern-.125emX}}
    
\begin{document}
\bstctlcite{IEEEexample:BSTcontrol}

\title{Boosting Performance Optimization with Interactive Data Movement Visualization}

\author{\IEEEauthorblockN{Philipp Schaad \orcidlink{0000-0002-8429-7803}}
\IEEEauthorblockA{
\textit{Department of Computer Science}\\
\textit{ETH Z\"urich}\\
philipp.schaad@inf.ethz.ch}
\and
\IEEEauthorblockN{Tal Ben-Nun \orcidlink{0000-0002-3657-6568}}
\IEEEauthorblockA{
\textit{Department of Computer Science}\\
\textit{ETH Z\"urich}\\
talbn@inf.ethz.ch}
\and
\IEEEauthorblockN{Torsten Hoefler \orcidlink{0000-0002-1333-9797}}
\IEEEauthorblockA{
\textit{Department of Computer Science}\\
\textit{ETH Z\"urich}\\
htor@inf.ethz.ch}
}

\maketitle

\begin{abstract}
Optimizing application performance in today's hardware architecture landscape is an important, but increasingly complex task, often requiring detailed performance analyses.
In particular, data movement and reuse play a crucial role in optimization and are often hard to improve without detailed program inspection.
Performance visualizations can assist in the diagnosis of performance problems, but generally rely on data gathered through lengthy program executions.
In this paper, we present a performance visualization geared towards analyzing data movement and reuse to inform impactful optimization decisions, without requiring program execution.
We propose an approach that combines static dataflow analysis with parameterized program simulations to analyze both global data movement and fine-grained data access and reuse behavior, and visualize insights in-situ on the program representation.
Case studies analyzing and optimizing real-world applications demonstrate our tool's effectiveness in guiding optimization decisions and making the performance tuning process more interactive.
\end{abstract}


\section{Introduction}\label{s:introduction}
With the end of Moore's Law~\cite{Moore1998} and Dennard scaling~\cite{Dennard1974}, performance optimization in modern high performance computing applications is more important than ever.
However, the complex and multifaceted nature of a program's performance in modern HPC environments necessitates conducting a careful and extensive performance analysis to make informed decisions about performance optimization steps.
Due to the increasing transistor efficiency on chips, but the relatively constant cost of transferring data, a particularly important metric for performance analysis and optimization is the amount of data movement, with the goal being to exploit data locality and reuse as much as possible~\cite{padal, Unat2017}.

To successfully analyze the performance of complex applications, engineers employ analysis tools such as profilers.
Further instrumenting code to read data from timers or hardware counters can help gather information about fine-grained program behavior.
These methods generate a wealth of data, which in itself can be difficult to understand without the help of performance visualization tools that aggregate and present it in an intuitive and understandable way.
However, gathering the raw data generally requires running the entire application, which in many cases can take prohibitively long and slow down the often iterative optimization process.

In this work, we provide a performance visualization technique that gathers and shows measurements based on simulation and static dataflow analysis to assist engineers in analyzing and optimizing the data movement and reuse behavior of their application, without requiring program execution.
We employ a graphical program representation that naturally exposes data flow, to build intuitive in-situ visualization overlays on the program's dataflow graph.
By mapping performance metrics directly onto the program, the cognitive load required for the attribution of observations to the original code, and the time required to perform a root-cause analysis are significantly reduced.

To reduce the need for costly, full-scale executions of an application, we propose an approach that simulates individual, parameterized program parts, to estimate an application's data access and reuse behavior.
By reducing the wait time for performance data from minutes or hours to a fraction of a second, this approach enables a more interactive performance optimization process.

To facilitate program analysis on multiple granularities, we construct our visualization using a two-level approach.
A coarse view mode provides a global understanding of the program and aims to support fast and reliable data movement bottleneck detection.
This is achieved by analyzing static program information such as logical data movement volumes and arithmetic operation counts, and visualizing the resulting metrics using in-situ overlays on a graphical program representation.
This view mode facilitates the analysis of the overall algorithmic design and data movement or communication scheme, and enables scalability analyses with rapid feedback through the use of symbolic analysis.

A second, fine-grained view mode allows for close-up analysis of data locality and reuse behavior in individual program parts.
This view mode connects statically obtained \emph{logical} data movement to \emph{physical} data movement, by simulating a program's data access patterns on a simple hardware model.
The view mode exposes this information alongside visualizations of temporal and spatial data locality, physical data layouts, and cache miss estimations.

We implement our visualization as a Visual Studio Code~\cite{vscode} extension and demonstrate its use in two realistic scenarios.
By analyzing and optimizing the encoder layer for a BERT transformer deep neural network~\cite{Devlin2019}, and a horizontal diffusion weather stencil, we demonstrate analysis of both global data movement and close-up memory layout and reuse.

In summary, this paper makes the following contributions:
\begin{itemize}
    \item Global data movement visualization for performance analysis.
    \item Approach to data movement estimation using parameterized, small-scale data access simulations.
    \item Visualization methods for hardware model-augmented data locality and reuse analysis.
\end{itemize}

\section{Background}\label{s:background}
The primary goal of performance engineering is to make an application perform more efficiently, typically by reducing its runtime.
For any reasonably large program, achieving this goal is a complex process.
An application's performance depends on a large number of factors, both from inside the program source code, but also external factors, such as the hardware or what data the program operates on.

To undertake effective optimizations, a performance engineer first needs to understand the performance characteristics of the application with a detailed analysis.
Only then can they perform educated optimization steps to improve particular aspects of the measured performance.
Given the complex interplay between different system factors, an application's performance then typically has to be reassessed before further optimizations can be performed, since even small changes and optimizations can have large effects on the previously measured performance.

This sequence of performance analyses followed by optimization steps is repeated until a program's performance targets are met.
Given the iterative nature of this procedure, shortening either the analysis or optimization step can create a large speedup in the overall performance engineering process.
With this work, we focus on reducing the time spent in the analysis step of this process.

\subsection{Performance Analysis}\label{ss:perf-analysis}
Model-driven performance tuning~\cite{Hoefler2011} suggests a top-down workflow during the analysis process.
Engineers initially identify theoretical limits to the achievable performance, gauging the proximity to the obtainable optimum.
They further identify input parameters, program parts (or \emph{kernels}), and communication or data movement patterns that impact the system's performance.
This information helps them focus further close-up analyses and optimization efforts with more fine-grained metrics.

Engineers have a variety of tools at their disposal to acquire such performance metrics.
They can choose to manually instrument their code to generate timing information, use hardware counters~\cite{Terpstra2010}, or run an application with detailed, and often hardware-specific profiling tools like NVIDIA's \texttt{nvprof}~\cite{nvprof}, Intel's \texttt{VTune}~\cite{intelVtune}, or \texttt{gprof}~\cite{Graham2004}.
The large amount of performance data generated by these methods is typically challenging to understand and needs to be carefully sifted through to extract useful information~\cite{Cornelissen2007}.

Performance visualization tools can help simplify the process of interpreting the wealth of generated performance data.
In particular, they present the data in a clear and aggregated manner, often highlighting a specific aspect of a program's performance.
This enables engineers to rapidly pinpoint what particular performance problems can be attributed to.

\section{Productive Performance Visualization}\label{s:architecture}
To support a top-down performance analysis process, we equip our visualization with two separate view modes, each specialized towards a specific set of tasks in different stages of the analysis workflow.
A global view is provided to assist in building an overview of the application and to perform an analysis of coarse program structures, such as communication and data movement patterns or the overall algorithmic design.
Additionally, the global view allows engineers to quickly identify program kernels or input parameters with high performance impact.

A second, local view specializes on fine-grained performance analysis of specific, smaller program parts.
The main task of this view is to highlight more detailed performance metrics that help inform specific tuning decisions.
Both view modes are described in detail in Sections \ref{s:global-view} and \ref{s:local-view}.

To best support engineers in a productive optimization and analysis process, and to facilitate a top-down information seeking behavior, we stipulate that our performance visualization should adhere to the following principles:

\begin{itemize}
    \item \textbf{Performance:} The visualization must primarily show factors that are important for performance optimization.
    \item \textbf{Minimality:} To avoid clutter, anything that is not strictly necessary for the task at hand, should not be shown.
    \item \textbf{Attribution:} It must remain clear what system or program elements specific parts of the visualization depict and relate to.
    \item \textbf{Continuity:} Changes in the visualization should not happen unexpectedly, and visual differences should carry a clear meaning.
\end{itemize}

\begin{figure*}
	\centering
	\includegraphics[width=.88\linewidth]{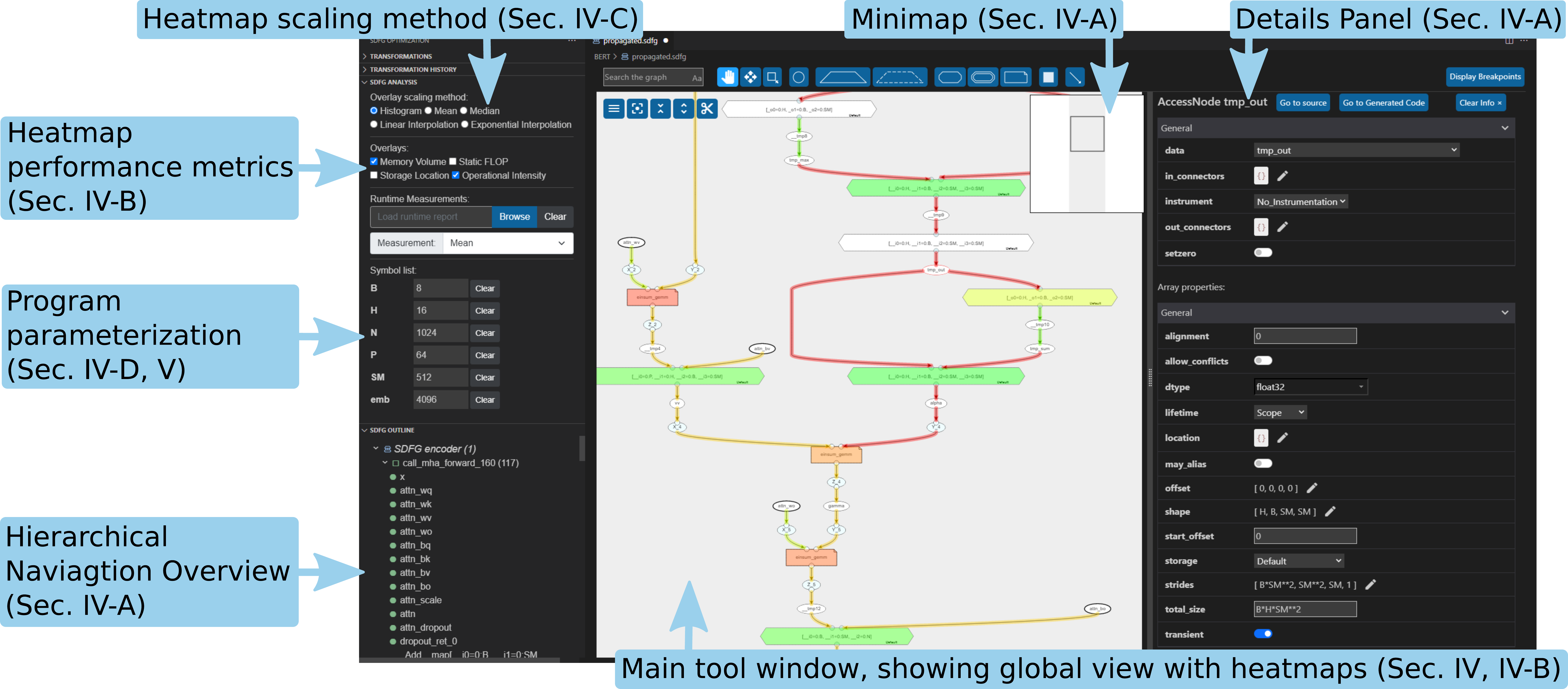}
	\caption{Screenshot of the visualization tool's main interface.}
	\label{fig:interface-overview}
	\vspace{-.5cm}
\end{figure*}

\subsection{Highlighting Data-Movement}\label{ss:highlighting-data-movement}
Based on the \textbf{performance} principle and the importance of data movement optimizations in modern HPC applications~\cite{padal, Unat2017}, we design our visualization centered around data movement, which can be analyzed separately from the computation and used to promote data locality and reuse.
To express this, we need to represent programs using a graph-based dataflow intermediate representation, which can be used to visualize performance metrics in-situ as overlays or augmentations directly in and on top of the program, simplifying the attribution of observations to the responsible program parts.
In such graphical dataflow representations, graph nodes represent computations and data containers, while directed edges between them represent data flowing through the program.

There are a number of graph-based dataflow programming languages or representations, like PROGRAPH~\cite{Matwin1985}, Stateful Dataflow Multigraphs (SDFGs)~\cite{Ben-Nun2019a}, or LabVIEW~\cite{Kodosky2020}, that can be used to express this aspect of a program.
For this work we use SDFGs to extract dataflow from general programs, because the accompanying data-centric parallel programming framework (DaCe) can compile programs from both Python and C programs~\cite{Calotoiu2021} into this intermediate representation, and they can directly be used for subsequent optimizations in DaCe.

\section{Global View: Data Movement Analysis}\label{s:global-view}
In our two-level approach to performance analysis, the global view is first tasked with providing a global comprehension of the program's performance characteristics.
The view is further responsible for exposing coarse data movement behavior, and highlighting the performance impact of individual program parts, to assist engineers in problem detection and diagnosis.
Lastly, the global view should facilitate the identification of input parameters which have a large impact on performance, by providing a fast-feedback scaling analysis.

\subsection{Global Comprehension}\label{ss:global-comprehension}
To give the engineer an overview of the global state of a program's behavior, this initial view provides mechanisms for exploring the entire application in its graphical representation.
When interacting with elements in this representation, the option is given to jump to the corresponding location in the original source code representation, helping to fulfill the \textbf{attribution} principle.

Maintaining an overview can be difficult for complex applications.
Particularly, as programs get larger, the number of visual elements needed to represent the application grows together with the number of code lines in the program's source representation.
To ensure legibility of large programs, it is imperative that the principle of \textbf{minimality} is followed.
Traditional code editors such as Visual Studio Code (VS Code)~\cite{vscode}, Atom~\cite{atom}, and most integrated development environments address this by allowing for logical code regions to be folded into a single line.
The SDFGs used to extract our dataflow graph representation are constructed in a hierarchical manner.
Individual program parts, or subgraphs, form logical regions similar to their counterparts in other high-level languages (e.g. loop contents and subroutines).
We exploit this to allow entire subgraphs to be folded and hidden, instead representing them with a single graph element that summarizes their content until they are deliberately expanded again.

To further improve legibility in large applications, more detailed visual elements are gradually hidden as the user zooms further out, similarly to Google Maps~\cite{gmaps}, dynamically pulling focus towards the more coarse-grained structure of the application.
Additionally, the graphical representation contains only information strictly necessary to analyze the functional behavior and dataflow of an application.
Any additional information like data types, sizes, and alignment are hidden away and appear on-demand in a separate details panel or in tooltips, shown only when the user interacts with the corresponding visual object.
As with traditional source code, the graphical representation can be searched to find specific elements, and it further allows for some types of elements to be filtered out and hidden from view.

The user can employ the type of pan-and-zoom navigation common in mapping software to explore the graphical program representation.
To assist in navigation in accordance with the principle of \textbf{continuity}, two separate overviews help maintain situational awareness.
A minimap in the top right corner of the visualization shows the current program in its entirety, with a box drawing the current viewport in relation to the graph.
A second, outline overview shows a hierarchical view of the graph, enabling quick navigation to a specific graph element by selecting it from this list.
This type of overview in combination with pan-and-zoom navigation helps in analyzing large program graphs, and has been shown to be a very efficient mode of exploration that is perceived as enjoyable by users~\cite{Nekrasovski2006}.
Any automatic navigation through interaction with one of the overviews is further animated as a slowed down motion of the viewport to maintain continuity and avoid disorientation.

An overview of the interface can be seen in Fig.~\ref{fig:interface-overview}, showing the global analysis view on the graphical program representation, embedded in the popular code editor VS Code.
A short supplementary video\footnote{Video `Program Navigation' (\url{https://youtu.be/anMPJ28dOO8})} demonstrates program navigation and exploration in the global view.

\subsection{Problem Detection and Diagnosis with Heatmaps}\label{ss:heatmap-problem-detection}
To facilitate efficient problem detection and diagnosis, the global view enables reasoning about the performance of the program as a whole.
Given the importance of reducing data movement when improving performance, it is crucial that optimizations focus on maximizing data reuse.
This can be achieved by coalescing computations that rely on the same data, which allows for better utilization of caches, in turn reducing the amount of data that needs to be read from or written to main memory.
In doing so, one increases the arithmetic intensity, i.e., the number of arithmetic operations performed per transferred data byte.

The amount of data being accessed by or moved between individual operations in the program is statically determined when SDFGs are generated in a dataflow extraction procedure~\cite{Calotoiu2021}.
We can use this information to augment the graphical program representation, by showing a color-coded heatmap as an overlay directly on top of the program's dataflow graph.
This heatmap uses a green-yellow-red color spectrum to mark data movement edges with a color corresponding to the relative amount of data being moved, where green represents a low amount of data, and red represents higher volumes.

We extract information on arithmetic or operational intensity separately by parsing the abstract syntax tree of individual computations, counting the number of arithmetic operations.
This can be used to directly construct a heatmap, coloring nodes based on their arithmetic operations count, or be combined with data access information to color nodes based on their arithmetic intensity.

This form of color-coded overlays shown directly on top of the program structure has been found to be an effective tool for communicating additional information to software engineers~\cite{Harward2010}.
We exploit this technique by applying it to the graphical program representation, with the goal of reducing the cognitive load required to perform attribution of individual measurements to their corresponding program parts.
Fig.~\ref{fig:interface-overview} shows the global view of a program colored using both a logical memory movement volume and arithmetic intensity heatmap.

The proposed visualization is not directly tied to static analysis. Profiling data could orthogonally be used as metrics, which would be crucial for bottleneck analysis of data-dependent programs.

\subsection{Heatmap Coloring}\label{ss:heatmap-coloring}
In real-world applications, we observe large value ranges for performance metrics, with data movement volumes ranging from individual bytes, to multiple megabytes or gigabytes.
These magnitude changes, even within applications, make determining an individual value's heatmap color with any fixed scale impractical.
Color scales consequently have to be adaptive and account for varying value distributions.

Heatmaps are a popular tool in performance visualization systems~\cite{Landge2012, Bhatele2012, Moreta2007, Geimer2010}, and there are different approaches to handle this.
The Scalasca toolkit's visualization component, Cube, has a separate plugin with options for the user to change this scaling behavior by switching the interpolation method used for determining the scale between using linear or exponential interpolation from the minimum to maximum observed values~\cite{Saviankou2015, cubeAdvancedColorMapPlugin}.

\begin{figure}[t]
	\centering
	\includegraphics[width=\linewidth]{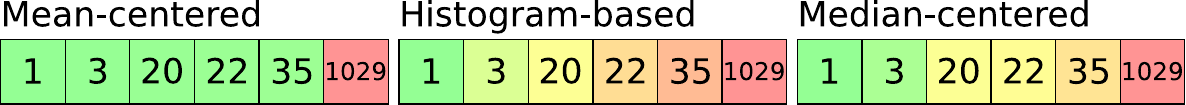}
	\caption{Heatmap scaling methods and their respective uses.}
	\label{fig:heatmap-scaling-demo}
\end{figure}

We employ three further approaches not based on interpolation, to serve three separate use cases.
The user can dynamically select the fitting one, updating the heatmap display.
The first two methods work by determining a center value $c$ for the color scale, and then setting the scale to run in the interval $[0, 2c]$.
The center value $c$ is determined by sorting all observed data values in increasing order, and then picking either the \emph{median} or \emph{mean} of all observations to form the center of the scale.
Observations above the maximum of the scale are clamped to $2c$.
In a third method, we group the observed values into histogram buckets and set the color scale to the interval $[0, n]$, where $n$ is the number of distinct buckets.
A value's color is then chosen based on the position of index $i$ of its corresponding histogram bucket on the scale.

Fig.~\ref{fig:heatmap-scaling-demo} demonstrates the use cases for each of these scaling methods, highlighting their behavioral differences.
Centering the scale around the mean (Fig.~\ref{fig:heatmap-scaling-demo}, left) is heavily influenced by outliers, making it ideal for detecting bottlenecks by giving them visually distinct colors from the remainder of the distribution.
Histogram-based scaling (Fig.~\ref{fig:heatmap-scaling-demo}, middle) distorts the scale to give each distinct observation a different color.
This is the most useful method for clearly highlighting the distribution of observed values, independently of the distances between observations.
Centering the scale around the median (Fig.~\ref{fig:heatmap-scaling-demo}, right) fills the gap between these two methods:
by being more outlier resistant than mean-centered scaling, outliers are less visually distinct, but by distorting the scale less than histogram-based scaling, this method is ideal for visually grouping values of similar magnitudes with similar colors.

The colors used to represent these scales are equally important, and some color schemes used in many visualization systems, such as rainbow maps (also known as \emph{jet} maps), are less than desirable and can be actively confusing~\cite{Moreland2016, Borland2007, Liu2018}.
To combat this, a popular alternative is a green-red spectrum, which leverages intuitive color associations of red$=$slow and green$=$fast.
This works well with sparse data sets, where individual values are well separated.
However, for data sets where distinction between individual values is important, and where values are closer relatively to one another, there is little visual separation between individual data points.

We address this by introducing the additional color yellow in the center to increase this separation while keeping the clear color ordering from fast to slow:

\vspace{-.05cm}
\begin{center}
    \includegraphics[width=\linewidth, height=.4cm]{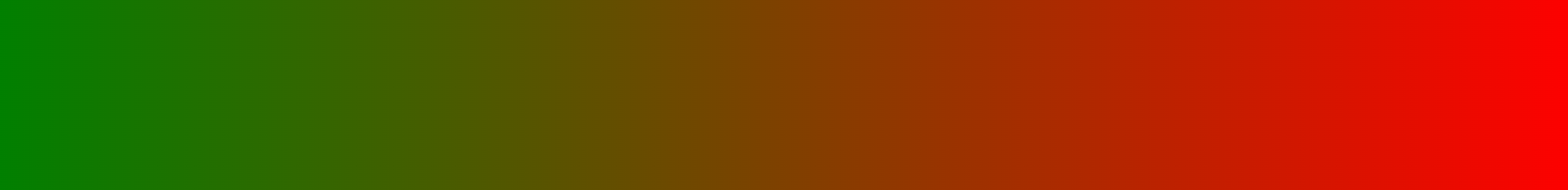}
\end{center}
\vspace{-.82cm}    
\begin{center}
    \includegraphics[width=\linewidth, height=.4cm]{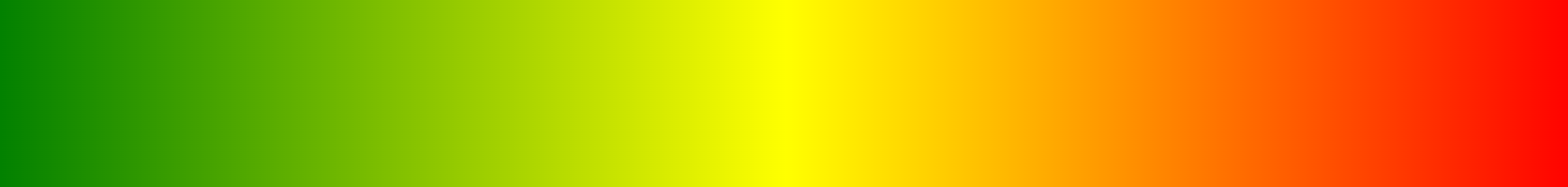}
\end{center}
\vspace{-.2cm}

To further account for color blindness, this color scale can be manually changed to fit the user's needs.

\subsection{Parametric Scaling Analysis}\label{ss:parametric-scaling-analysis}
In many applications, computations and program behavior are largely dependent on the input size of the data or other runtime parameters.
As such, many resulting arithmetic operations or data volume counts are expressed as symbolic expressions that depend on input data dimensions or parameters passed to the program at runtime.
The SDFG representation used in the global view is thus inherently \emph{parametric}, a fact that can be exploited to observe how program performance is affected by changes to input parameters, and by extension identify which parameters have a particularly large impact on program performance.

By allowing the user to define and change values for input parameters in a configuration panel, we can adapt the heatmap visualizations on the fly by re-evaluating symbolic expressions with the new values.
This facilitates an analysis of the program's scaling behavior with respect to input sizes or runtime parameters.
The user can visually follow how data movement volumes, operational intensity, or localized bottlenecks are affected by changes to the data sizes.
With this, we allow the user to interactively determine what input parameters are crucial factors in the program's performance, without requiring costly program executions.

The use of heatmaps, including parametric scaling, is further demonstrated in a short supplementary video\footnote{Video `In-Situ Overlays' (\url{https://youtu.be/4s996YtZvKk})}.

\section{Local View: Locality and Reuse Analysis}\label{s:local-view}
To perform close-up analyses of data locality and reuse behavior, we employ a small-scale simulation approach, which approximates values for the program's runtime behavior with respect to data accesses and layouts.
Observations of this behavior are conventionally obtained via profiling and program counter collection, or with costly hardware simulations.
This approach is particularly well suited for localized bottleneck analysis, avoiding executions of the entire application.
By instead solely simulating the relevant program kernel, this enables more interactive, fast-iterating optimizations.

Engineers can specify a region of the program where a closer investigation is desired, and provide sample values for input parameters that determine execution behavior, such as data dimensions.
This focuses the view on only the specified part of the graphical representation, and adapts the visualization to better facilitate this close-up analysis.

\subsection{Program Parameterization}\label{ss:parameterization}
With provided input parameters, we \emph{parameterize} the program view, augmenting graph elements that depend on input parameters to reflect the given concrete values.
Specifically, nodes representing data containers such as arrays are now expanded to reflect their parameterized size, showing each individual data element, to represent individual memory locations.

\begin{figure}
	\centering
	\includegraphics[width=.66\linewidth]{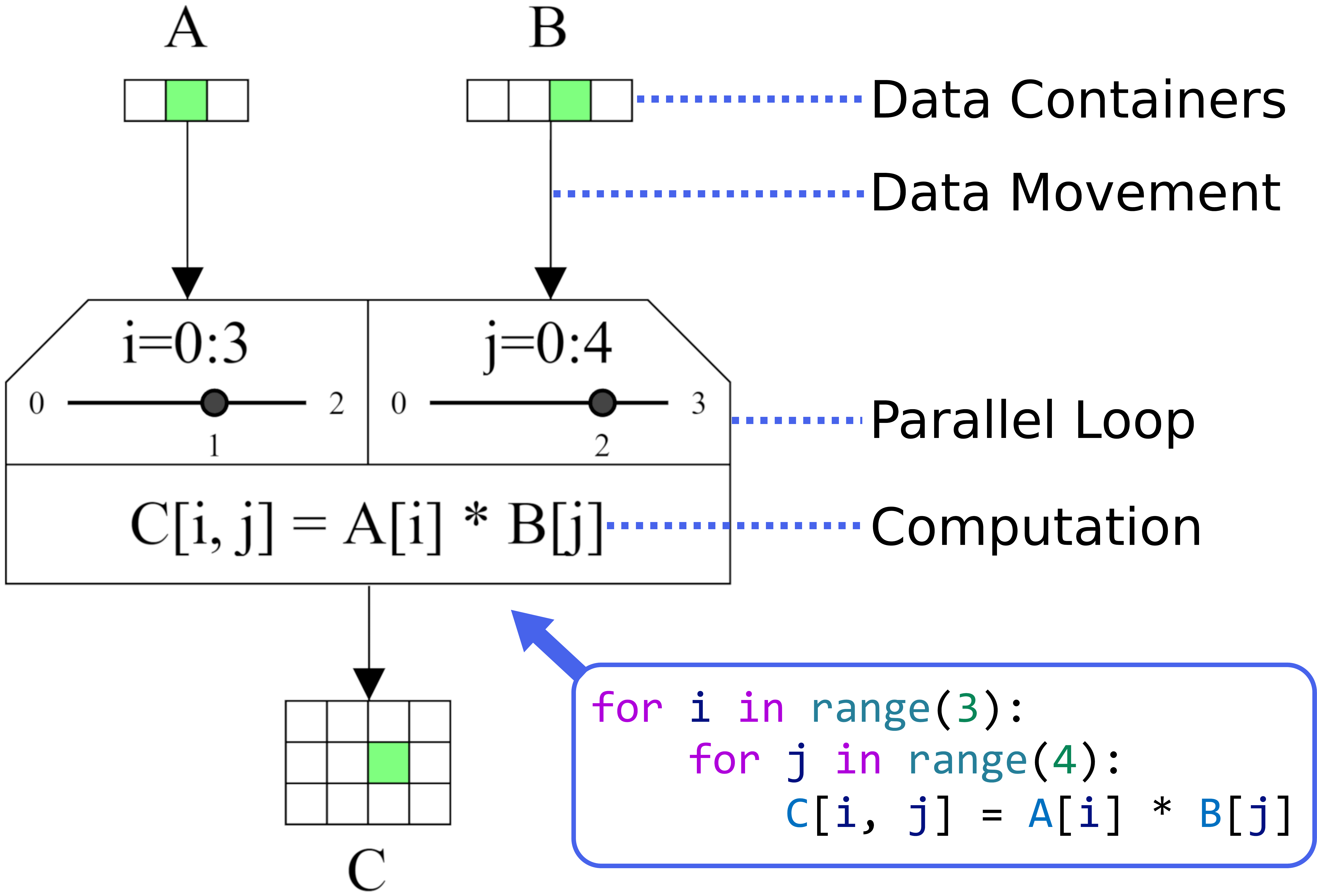}
	\caption{Parameterized outer vector product $C=A\otimes B$, for $A\in\mathbb{R}^3$, $B \in \mathbb{R}^4$, and $C \in \mathbb{R}^{3\times4}$. The sliders set to the loop parameters $i=1$ and $j=2$ highlight memory elements accessed with those parameters (green).}
	\label{fig:param-view-example}
\end{figure}

Parallel loops and concurrent regions with parametric bounds are also parameterized and have their bounds fixed.
These structures are shown as boxes with trapezoidal header bars, where everything inside the box represents the loop or concurrent region's contents, and the trapezoidal header bar shows loop parameters and their bounds.
The example in Fig.~\ref{fig:param-view-example} shows the outer product calculation of two vectors $A \in \mathbb{R}^{3}$ and $B \in \mathbb{R}^{4}$, where a parallel loop with an iteration space given by $i \in [0,2]$ and $j \in [0,3]$ contains the calculation $C[i,j] = A[i] * B[j]$ in its loop body.
All data containers show an individual tile for each element.
Additionally, each parameter to the parallel region is accompanied by a slider with which the engineer can set individual parameter values.
This consequently highlights all memory elements accessed inside the parallel region for that specific parameter combination, as demonstrated in Fig.~\ref{fig:param-view-example} for parameters $i=1$, $j=2$.

\subsection{Visualizing High-Dimensional Data}\label{ss:high-dim-data}
While it is intuitive to visualize one or two dimensional data on a 2D surface such as the user's screen, many programs deal with data containing more than two dimensions.
Shneiderman's taxonomy for visualization systems~\cite{Shneiderman1996} identified a number of ways to deal with this, usually involving filtering or slicing to only view specific sub-dimensions at a time.
While this makes it easier to visualize specific portions of the data and removes visual confusions, the burden of keeping a global picture of the remaining, hidden data is usually left to the user or has to be visualized separately.

\begin{figure*}[t!]
	\centering
	\begin{minipage}[b]{.32\linewidth}
		\begin{subfigure}[b]{\linewidth}
			\centering
			\includegraphics[width=.92\linewidth]{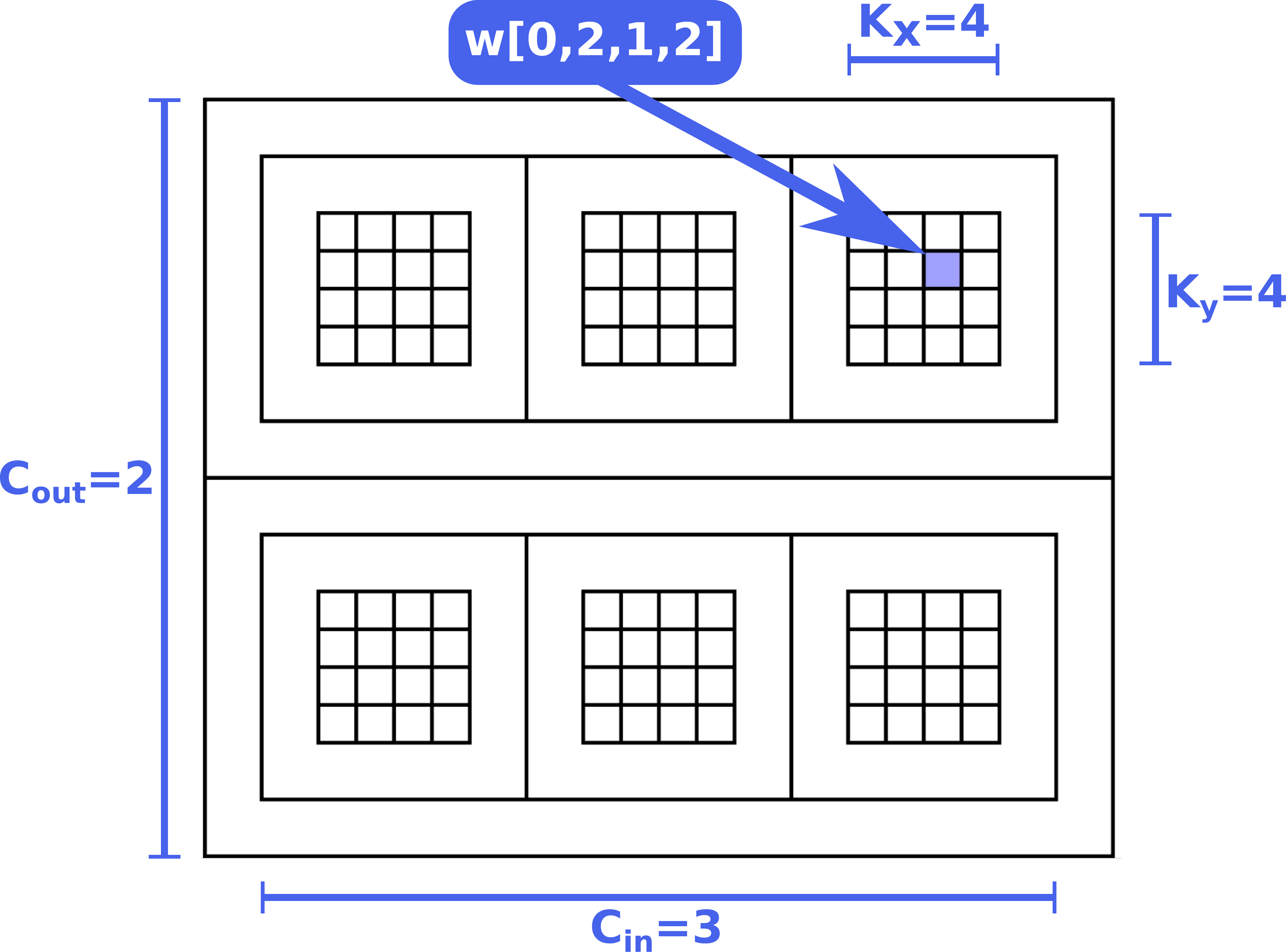}
			\vspace{1.4cm}
			\caption{Four-dimensional container for 3D convolution weights $w \in \mathbb{R}^{C_{out} \!\times\! C_{in} \!\times\! K_y \!\times\! K_x}$.}
			\label{fig:high-dim-data}
		\end{subfigure}\vspace{.2cm}
	\end{minipage}\hfill
	\begin{subfigure}[b]{.34\linewidth}
		\centering
		\includegraphics[width=\linewidth]{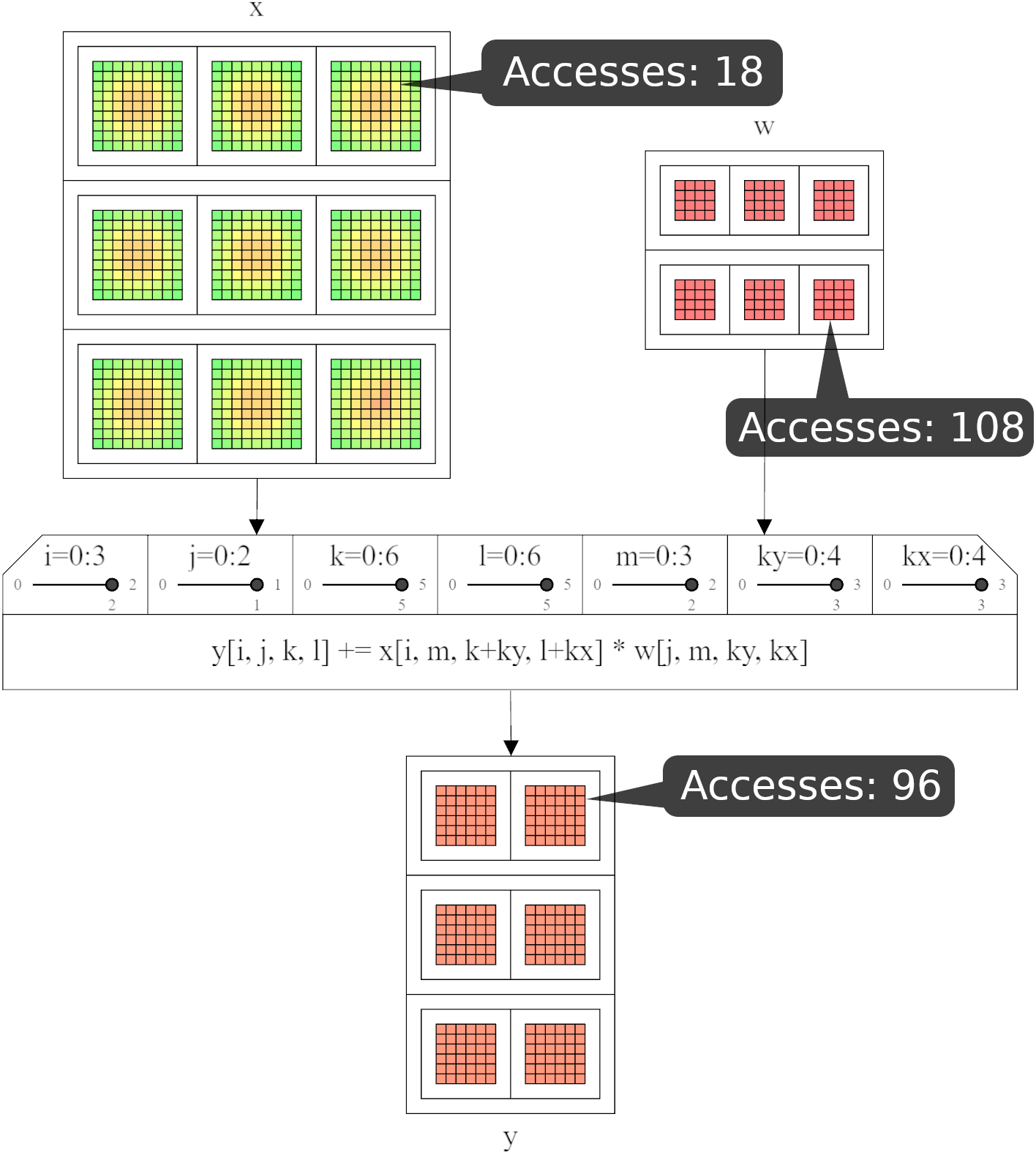}
		\caption{Distribution of the number of accesses in a 3D convolution without padding, which maps 3-channel, $9\!\times\!9$ inputs to 2-channel, $6\!\times\!6$ outputs.}
		\label{fig:access-pattern-example}
	\end{subfigure}\hfill
	\begin{subfigure}[b]{.31\linewidth}
		\centering
		\includegraphics[width=.85\linewidth]{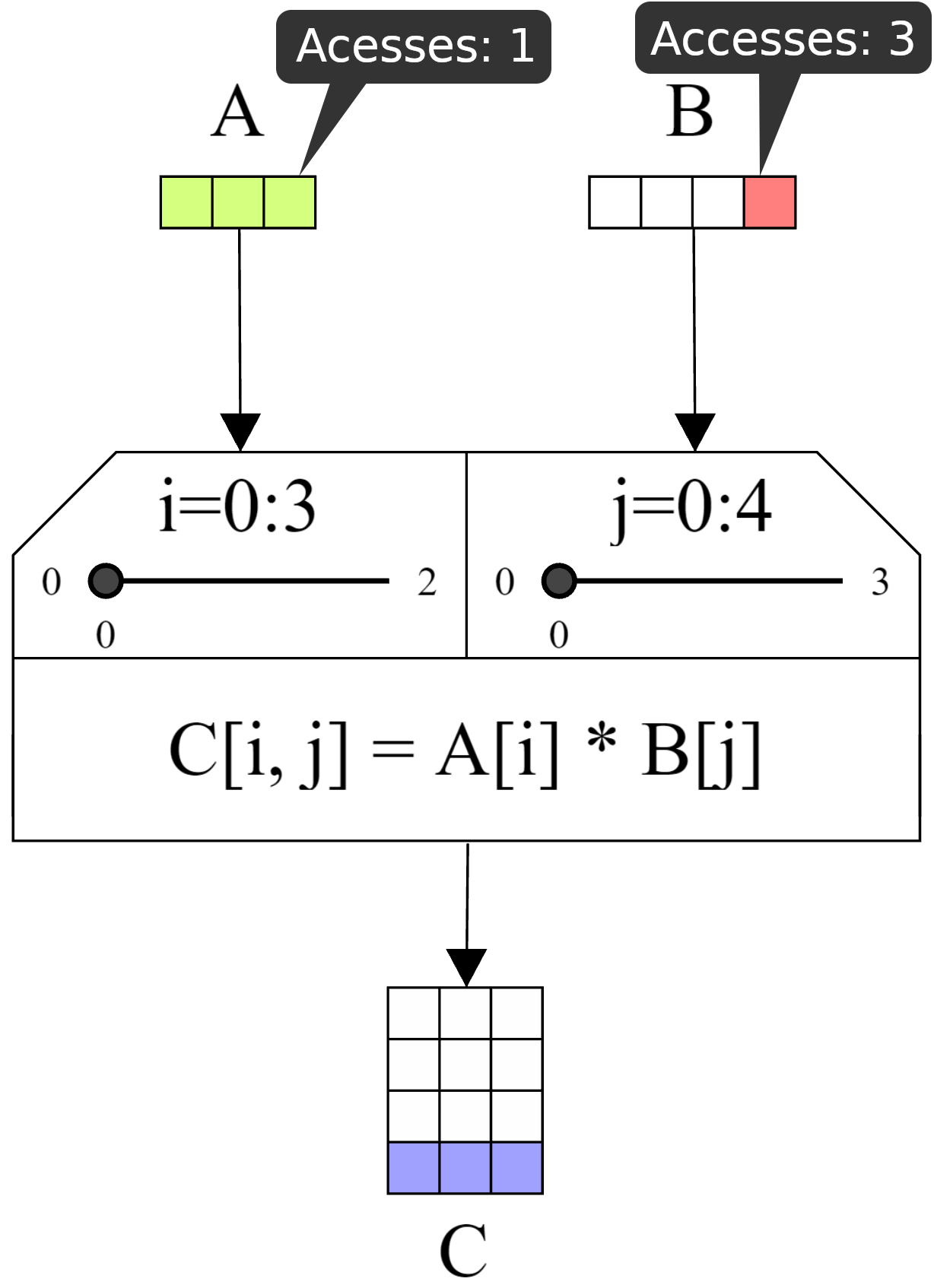}
		\caption{Showing related accesses to $A$ and $B$ for accesses to $C[3, 0]$, $C[3, 1]$, and $C[3, 2]$ in an outer vector product $C = A \otimes B$.}
		\label{fig:related-accesses-example}
	\end{subfigure}
	\caption{Multi-dimensional data containers, and screenshots of access pattern visualizations in the parameterized view.}
	\label{fig:param-view-heatmaps-demo}
\end{figure*}

We propose an alternative approach, in which we aim to visualize the entire data simultaneously using a hierarchical view that mimics how multidimensional arrays are abstracted in most high-level programming languages.
The two innermost dimensions are laid out in a 2D grid, and those are nested in alternating horizontal and vertical 1D grids for the remaining higher dimensions.
For one dimensional data containers, a simple 1D grid is used.
An example of this can be seen in Fig.~\ref{fig:high-dim-data}, where the weight tensor $w \in \mathbb{R}^{C_{out} \!\times\! C_{in} \!\times\! K_y \!\times\! K_x}$ for a 3D convolution is shown.
While this representation rapidly grows in space with more added dimensions, it works well in this parameterized setting, due to the small values expected for each individual dimension.

\subsection{Access Pattern Simulation}\label{ss:access-patterns}
With both data sizes and parallel region parameters specified, we can derive the exact access pattern for each data container in the graph.
To do this, we exploit the fact that programs translated to the SDFG intermediate representation carry an annotation of exactly what data subsets are being accessed by each computation in the form of a symbolic expression.
This expression can traditionally not be evaluated statically without dependent symbol values such as loop iteration variables.
In the parameterized graph, where parallel regions have their bounds fixed, we can perform an iteration space simulation to evaluate these symbolic expressions and derive the exact data accesses performed by each computation in the graph.
By extension, this determines the exact access pattern for each data container.

The resulting access pattern can be played back using a variable speed animation, which highlights the exact individual elements or memory locations in each data container accessed at that specific time-step of the simulation.
Alternatively, the time dimension can be flattened, summing up the number of accesses for each element in each data container, and showing the resulting distribution using a colored heatmap, where elements with a higher number of accesses are colored red, and lower numbers are represented in green.
The exact number of accesses can be viewed using a tooltip when hovering the corresponding data elements.
An example of the flattened time dimension using a heatmap can be seen in Fig.~\ref{fig:access-pattern-example}, where the access pattern of a 3D convolution is shown with tooltips superimposed.

\begin{figure*}[t!]
	\centering
	\begin{subfigure}[b]{.26\linewidth}
		\centering
		\includegraphics[width=\linewidth]{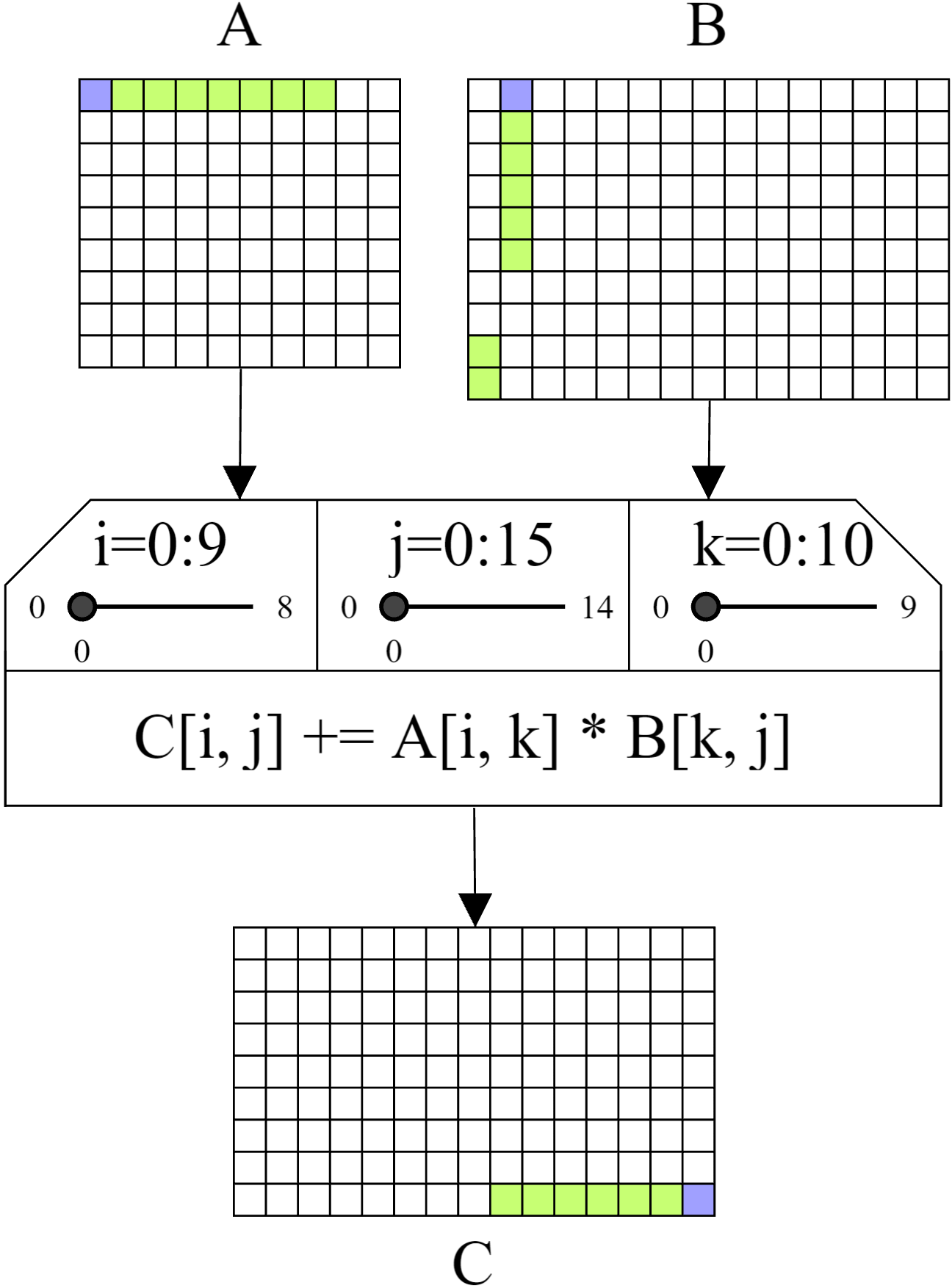}
		\caption{Visualizing data layouts by highlighting spatial locality in the form of cache lines.}
		\label{fig:cache-line-example}
	\end{subfigure}\hfill
	\begin{subfigure}[b]{.32\linewidth}
		\centering
		\includegraphics[width=.83\linewidth]{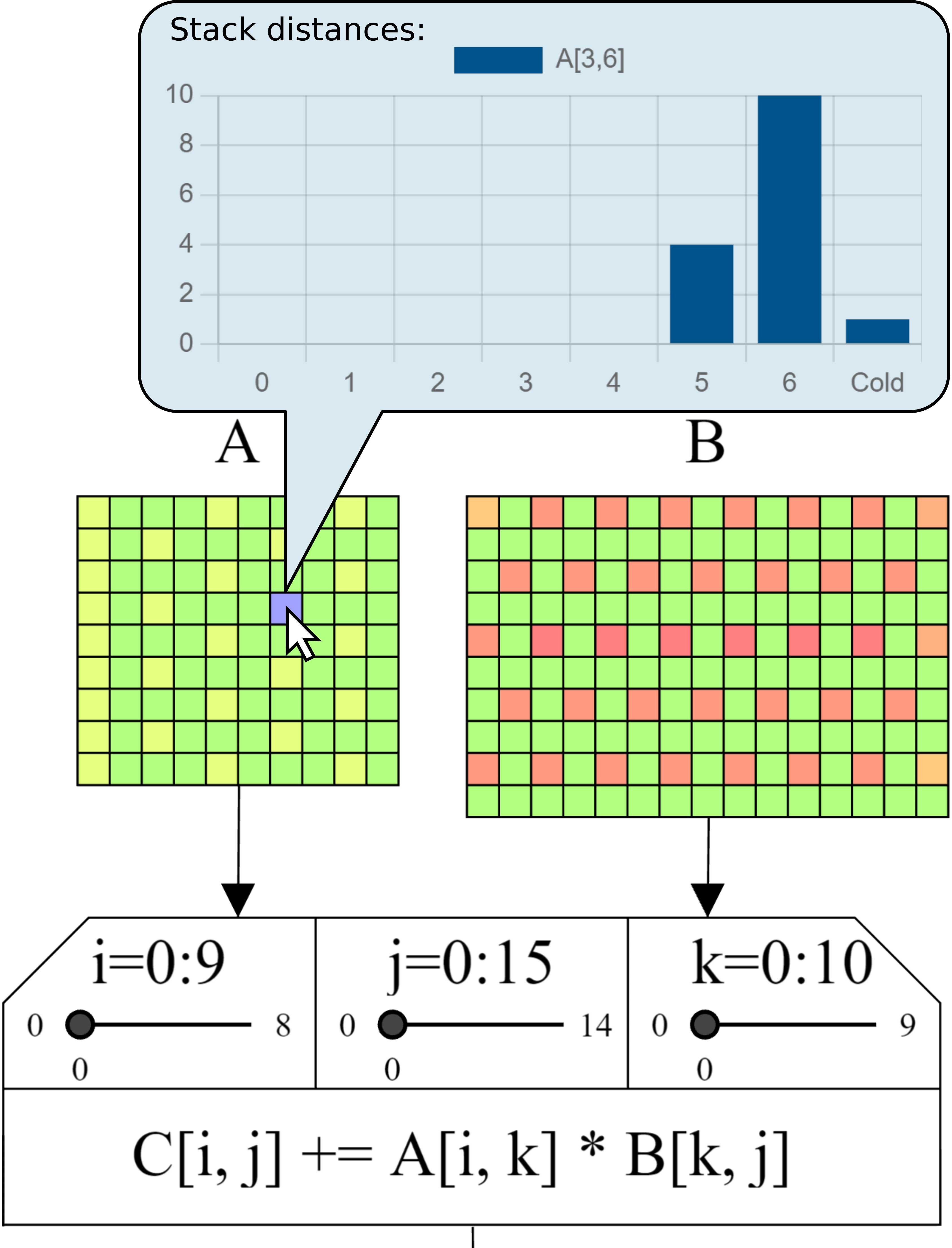}
		\caption{Distribution of median reuse distance per element, with a detailed histogram breakdown (top).}
		\label{fig:reuse-distance-example}
	\end{subfigure}\hfill
	\begin{subfigure}[b]{.39\linewidth}
		\centering
		\includegraphics[width=\linewidth]{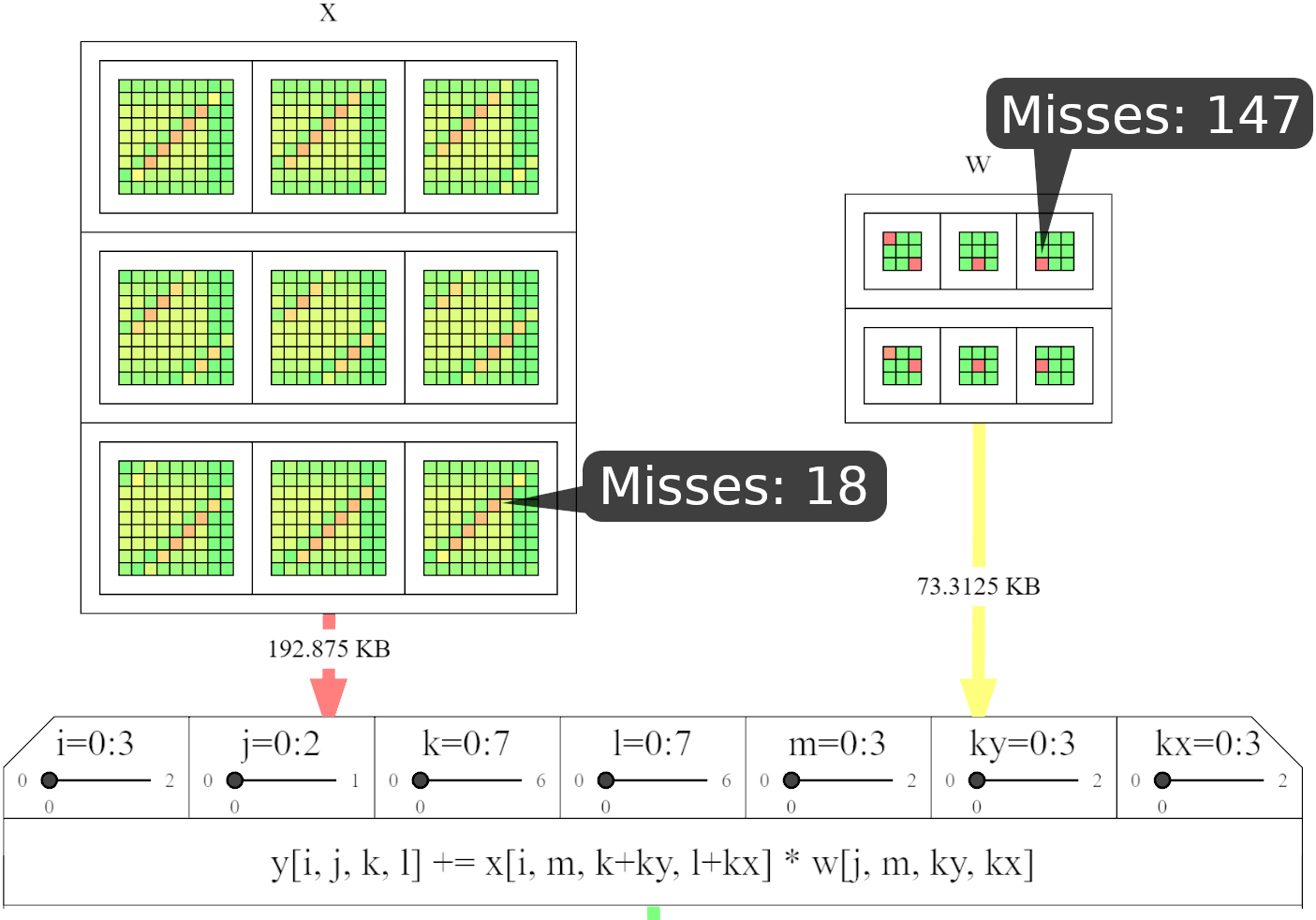}
		\vspace{.2cm}
		\caption{Estimated cache misses and physical data movement overlay based on reuse distances and cache line sizes.}
		\label{fig:physical-data-movement-example}
	\end{subfigure}
	\caption{Screenshots from our tool, visualizing physical data layouts, reuse distances, and estimated physical data movement.}
	\label{fig:reuse-distance-phys-data-movement-example}
\end{figure*}

The same information can be used to derive and visualize data accesses related to other accesses, based on whether they occur in the same computations.
For example, in the case of an outer product calculation $C[i,j] = A[i] * B[j]$, where $i \in [0, 2]$ and $j \in [0, 3]$, an access to $B[0]$ is associated to accesses of $C[i, 0]$ and $A[i]$, for all $i \in [0, 2]$.
The engineer can click one or more memory locations, stacking the number of related accesses to form a corresponding access heatmap, which helps analyze for potential replication or loop tiling opportunities.
An example of this can be seen in Fig.~\ref{fig:related-accesses-example}.

\subsection{Data Layout and Spatial Locality}\label{ss:data-layout}
Apart from the access pattern, the second important factor that determines how efficiently data locality is being leveraged, is the physical data layout.
This is usually opaque to the engineer and needs to be derived from the alignment, offsets, and padding used by the compiler or a specific data structure.
However, this information is crucial in determining how efficiently the cache is used to exploit \emph{spatial locality}.

To expose this information to the user, we provide an overlay that visualizes which elements are adjacent to one another in memory, by highlighting which data elements are pulled into the cache alongside a specified other element.
To determine this, an engineer must only provide the cache line (cache block) size in bytes for the target architecture.
The remaining information, like individual element sizes, alignment, offset, and padding, can all be extracted from the program's intermediate representation.
The user can then select memory elements, and the overlay highlights any other memory elements that fit into the same cache line, effectively exposing the physical data layout to the user and providing a guidance for exploiting spatial locality.

Fig.~\ref{fig:cache-line-example} shows the parameterized matrix multiplication of two matrices $A \in \mathbb{R}^{9\!\times\!10}$ and $B \in \mathbb{R}^{10\!\times\!15}$, where each value is 4 bytes in size and the cache line size is set to 64 bytes.
By selecting $A[0, 0]$, $B[0, 1]$, and $C[8, 14]$, the visualization highlights all elements pulled into cache together with these accesses in green, revealing that $A$ and $C$ are stored in row-major format, while $B$ is stored in column-major ordering.

\subsection{Stack Distance and Temporal Locality}\label{ss:reuse-distance}
The obtained exact data access patterns can further be used to expose \emph{temporal locality}.
To do this, we calculate a metric called the \textbf{stack distance} for each data element, which is defined as the number of accesses to \emph{unique} addresses made since the last reference to the requested data element~\cite{coffman1973operating}.
We use the stack distance at a cache line granularity, meaning that for each reference to a data element, all other elements in the same cache line are also referenced, resetting their stack distance to zero.
If an element has not been referenced yet, its stack distance is set to infinity.

The distribution of stack distances for each individual memory element can be visualized in-situ using a heatmap.
The engineer can choose whether the heatmap should visualize the distribution of the minimum, maximum, or median stack distances for each element.
To analyze this data at a finer granularity, an additional histogram with all the calculated stack distances over time is plotted in the details panel when an individual memory element or container is selected.
Fig.~\ref{fig:reuse-distance-example} shows a heatmap for median reuse distances on the input matrices to the previously used matrix multiplication, using a cache line size of 32 bytes.
Selecting the element $A[3, 6]$ plots a histogram that reveals the exact distribution, and shows that this element was accessed once without having previously been moved to the cache, by listing one cold miss.

\subsection{Cache Misses and Physical Data Movement}\label{ss:phys-data-movement}
With a now complete prediction of the program's data access behavior, including both temporal and spatial locality, we can construct a rough prediction on the number of cache hits and misses~\cite{Beyls2001}.
This in turn can be used to refine the abstract, logical memory movement volume shown on memory movement edges in the global, parametric view.
To count the number of predicted cache misses, there are three types of misses that should be considered.

\paragraph{\textbf{Cold miss}}\label{p:cold-misses}
A cold miss happens when a memory address is accessed for the first time without having been referenced before via spatial locality (i.e., in the same cache line as a different referenced address).
We count a cold miss for every access where the stack distance is infinite.

\paragraph{\textbf{Capacity miss}}\label{p:capacity-misses}
A capacity miss is encountered when a memory address is accessed after it has been evicted from the cache because of the eviction strategy after the cache or associated cache set has become full.
Assuming an LRU or LRU-derived eviction strategy, we count a capacity miss for every access where the stack distance is above a certain threshold value.
To model different cache sizes or degrees of cache associativity, the user can modify this threshold on-the-fly through the user interface.
It can further be used to adjust for the fact that the simulated data sizes are not equal to the expected data sizes in the target environment, which would make calculations based on the total cache size unrealistic.

\begin{table*}[t]
    \centering
    \caption{Case Study Benchmark Results}
    \begin{tabular}{llrrrrrrrr}
        \toprule
        \textbf{Application} & & \multicolumn{6}{c}{\textbf{System}} \\
        \cmidrule(lr){3-8}
        & & \multicolumn{2}{c}{Piz Daint (Supercomputer)} & \multicolumn{2}{c}{High-Performance Workstation} & \multicolumn{2}{c}{Consumer Hardware} \\
        \cmidrule(lr){3-4}
        \cmidrule(lr){5-6}
        \cmidrule(lr){7-8}
        & & Time \footnotesize{[ms]} & Speedup & Time \footnotesize{[ms]} & Speedup & Time \footnotesize{[ms]} & Speedup \\
        \midrule
        \textbf{BERT encoder} & Baseline & $8254.13$ & $1.0 \times$ & $13670.66$ & $1.0 \times$ & $8959.78$ & $1.0 \times$ \\
        & 1st set of loop fusions & $2273.37$ & $3.6 \times$ & $2443.13$ & $5.6 \times$ & $1427.03$ & $6.3 \times$ \\
        & 2nd set of loop fusions & $1163.36$ & $7.1 \times$ & $452.54$ & $30.2 \times$ & $336.84$ & $26.6 \times$ \\
        \midrule
        \textbf{Horizontal diffusion} & Baseline & $667.54$ & $1.0 \times$ & $449.63$ & $1.0 \times$ & $358.39$ & $1.0 \times$ \\
        & Best NPBench CPU result & $31.65$ & $21.1 \times$ & $18.43$ & $24.4 \times$ & $41.33$ & $8.7 \times$ \\
        & Hand-tuned using our tool & $4.41$ & $151.4 \times$ & $3.26$ & $138.0 \times$ & $7.00$ & $51.2 \times$ \\
        \bottomrule
    \end{tabular}
    \label{tab:case-studies-results}
    \vspace{-.5cm}
\end{table*}

\paragraph{\textbf{Conflict miss}}\label{p:conflict-misses}
A conflict miss occurs when a memory address is accessed after it has been evicted due to a conflict, meaning a different element and its cache line was mapped to the same location in the cache.
This type of miss can only appear in set-associative or direct mapped caches.
Because the physical addresses of data containers depend on runtime conditions and their respective sizes, counting this type of cache miss in a small-scale, parameterized setting may introduce significant complexity in interpreting and generalizing the results.
To combat this, we assume a \emph{fully-associative} cache and do not count conflict misses.
McKinley and Temam~\cite{McKinley1999}, and Beyls and D'Hollander~\cite{Beyls2001} show that this gives a good prediction for the total number of cache misses in low set-associative or direct mapped caches, since most cache misses can be attributed to capacity, and only a minority of misses are due to conflicts.
\smallskip

The resulting distribution and number of cache misses can be directly visualized using the same heatmap technique, or can be used to derive a rough prediction for the amount of physical data that needs to be moved to and from main memory.
We can obtain this estimate for each data movement edge by multiplying the number of misses in both the edge's source and destination nodes, with the number of bytes per cache line.
The resulting value can be used to refine the heatmap on the data movement overlay, as shown in Fig.~\ref{fig:physical-data-movement-example}, where the number of estimated cache misses and data movement volume are visualized on top of the input and weight tensors to a 3D convolution, using a cache line size of 64 bytes and 8 byte data values.
This enables informed decisions about where local replication or changes to the data layout could be beneficial, or where to apply techniques like loop tiling, fusion, or reordering.

\section{Case Studies}\label{s:case-study}
To demonstrate the use of our visualization in a realistic scenario, we walk through the performance analysis and optimization
procedure of two real-world HPC applications.
We evaluate each application on three systems:
\begin{itemize}
    \item The Swiss National Supercomputing Center's \emph{Piz Daint} supercomputer, which is a cluster of 5,704 Cray X50 nodes with a 12-core Intel Xeon E5-2690 v3 CPU at 2.6 GHz and 64 GB of RAM each. We run each application on a single Cray X50 node.
    \item A high-performance workstation with two 16-core Intel Xeon Gold 6130 CPUs at 2.1 GHz and 1.5 TB of RAM.
    \item A consumer-grade system with a 10-core Intel i9-7900X CPU at 4.5 GHz and 32 GB of RAM.
\end{itemize}
For all experiments we use Python 3.8 with DaCe version 0.13 and GCC 8.3.
Each experiment is run 100 times and we report the median runtime.

\subsection{BERT Transformer}\label{ss:bert-case-study}
Machine learning is particularly data intensive, making data movement costs harder to analyze.
With this case study, we demonstrate how a program's global data movement scheme can successfully be optimized with the help of information exposed through our visualization's global view.
For this purpose we select the encoder layer from the natural language model BERT~\cite{Devlin2019}.
This type of Transformer~\cite{Vaswani2017a} is a widely used neural network, where even pre-trained models take hours to tune in large-scale distributed environments.
We use a NumPy~\cite{numpy} implementation of this application as our baseline, using Intel MKL~\cite{mkl} to accelerate linear algebra operations.
The input parameters are selected to match the ones used in the original BERT publication~\cite{Devlin2019} ($\textsc{BERT}_\textsc{LARGE}$),
with a batch size $B=8$, $H=16$ attention heads, an embedding size $I=1024$, an input/output sequence length $SM=512$,
an intermediate size $emb=4096$,
and a projection size $P=\frac{I}{H}=64$.

\begin{figure}
	\centering
	\includegraphics[width=\linewidth]{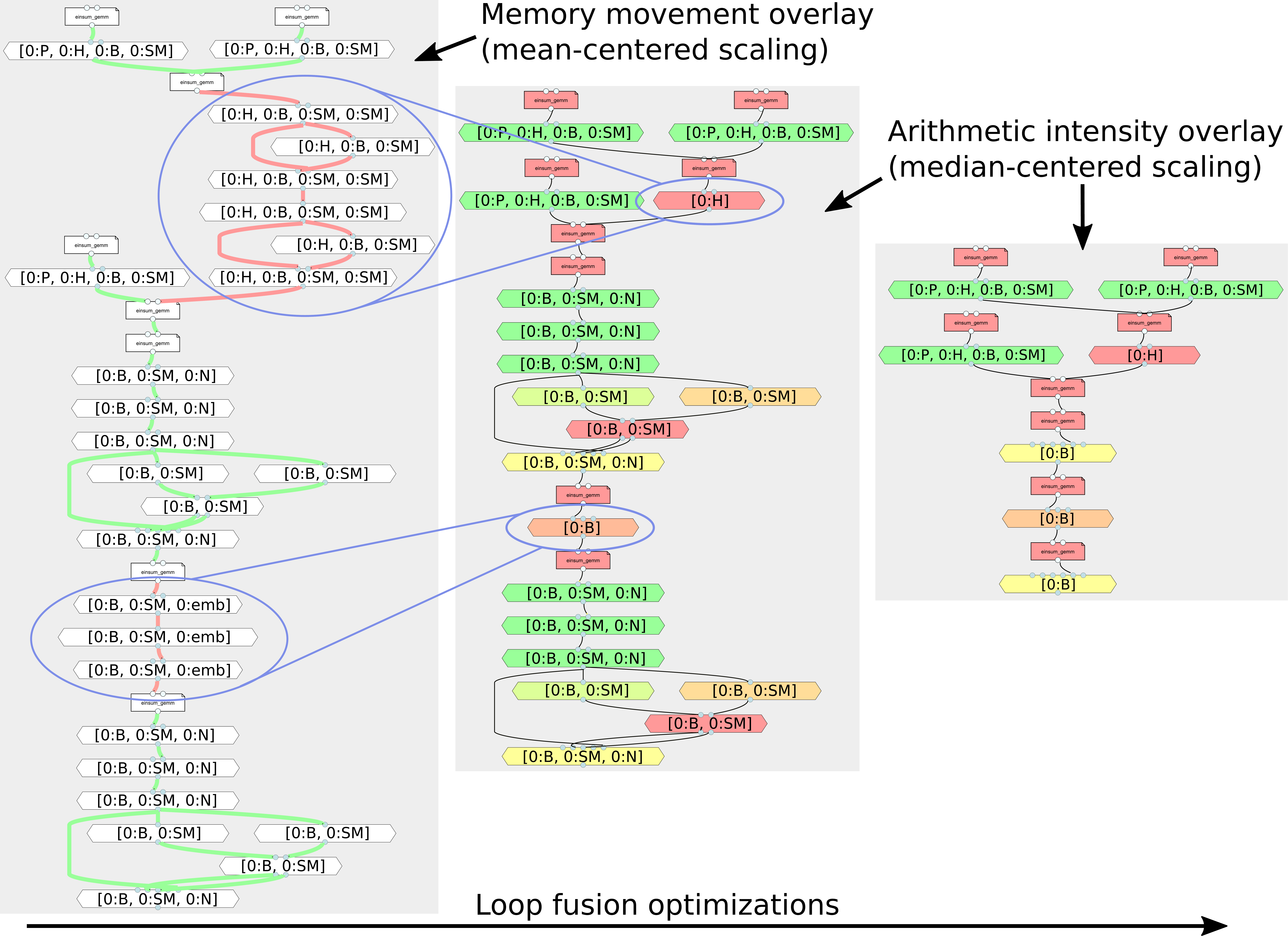}
	\vspace{-.5cm}
	\caption{Annotated snapshots of our global view, showing the BERT encoder layer at different stages of optimization.}
	\label{fig:bert-case-study}
\end{figure}

\begin{figure*}
	\centering
	\includegraphics[width=\linewidth]{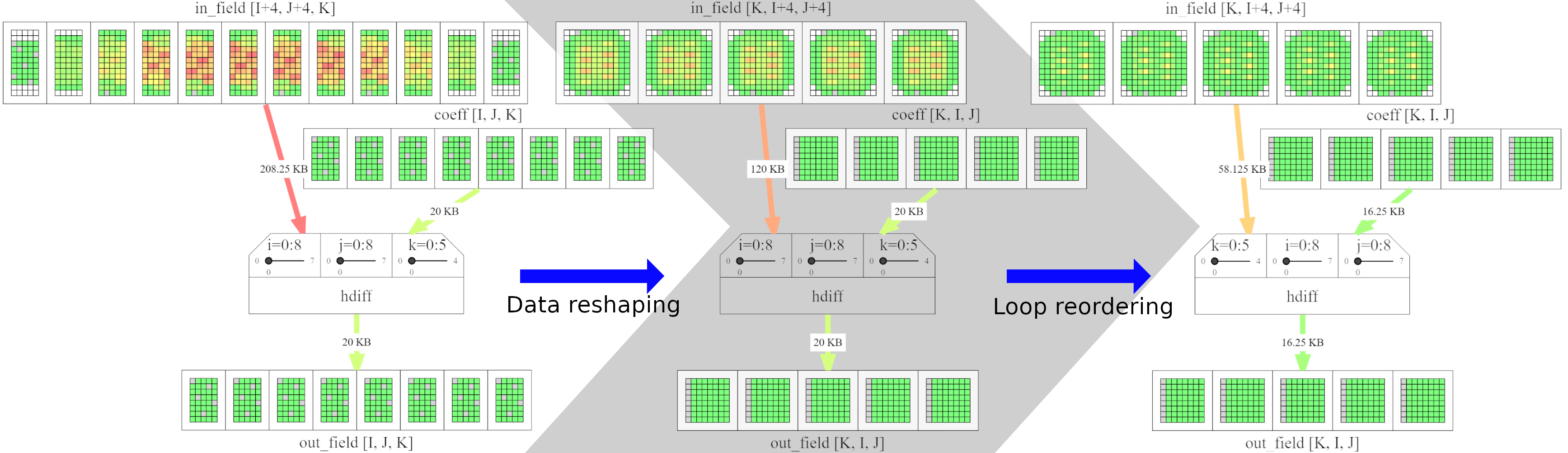}
	\caption{Screenshots of the local view, showing the number of cache misses and physical data movement for horizontal diffusion through the optimization process.}
	\label{fig:hdiff-progression}
	\vspace{-.2cm}	
\end{figure*}

Despite the large graph generated by this program, turning on heatmap overlays immediately helps identify some problems.
The logical data movement heatmap with scaling around the mean (shown in Fig.~\ref{fig:bert-case-study}, left) reveals two distinct series of edges highlighted in red.
This indicates that large amounts of data are being moved between individual graph nodes.
Clicking those nodes reveals that they represent parallel loops over similar loop bounds,
which we can combine via \emph{loop fusion}, removing the data movement between them.
This results in a new graph where these high-volume data movement edges are not present anymore (Fig.~\ref{fig:bert-case-study}, center).
This optimization already provides a significant speedup of between $3.6 \times$ and $6.3 \times$ over the baseline implementation,
as shown in \autoref{tab:case-studies-results}.

Using the arithmetic intensity overlay with median-centered scaling, as shown in Fig.~\ref{fig:bert-case-study} (center),
we can see a few computation nodes with a relatively low arithmetic intensity highlighted in green.
By clicking these nodes, the details panel again reveals that they represent parallel loops which can be fused together,
resulting in the much smaller program graph on the right side of Fig.~\ref{fig:bert-case-study},
where the number of computation nodes with low arithmetic intensity is visibly reduced.

\begin{figure*}
	\centering
	\begin{subfigure}[b]{.28\linewidth}
		\centering
		\includegraphics[width=\linewidth]{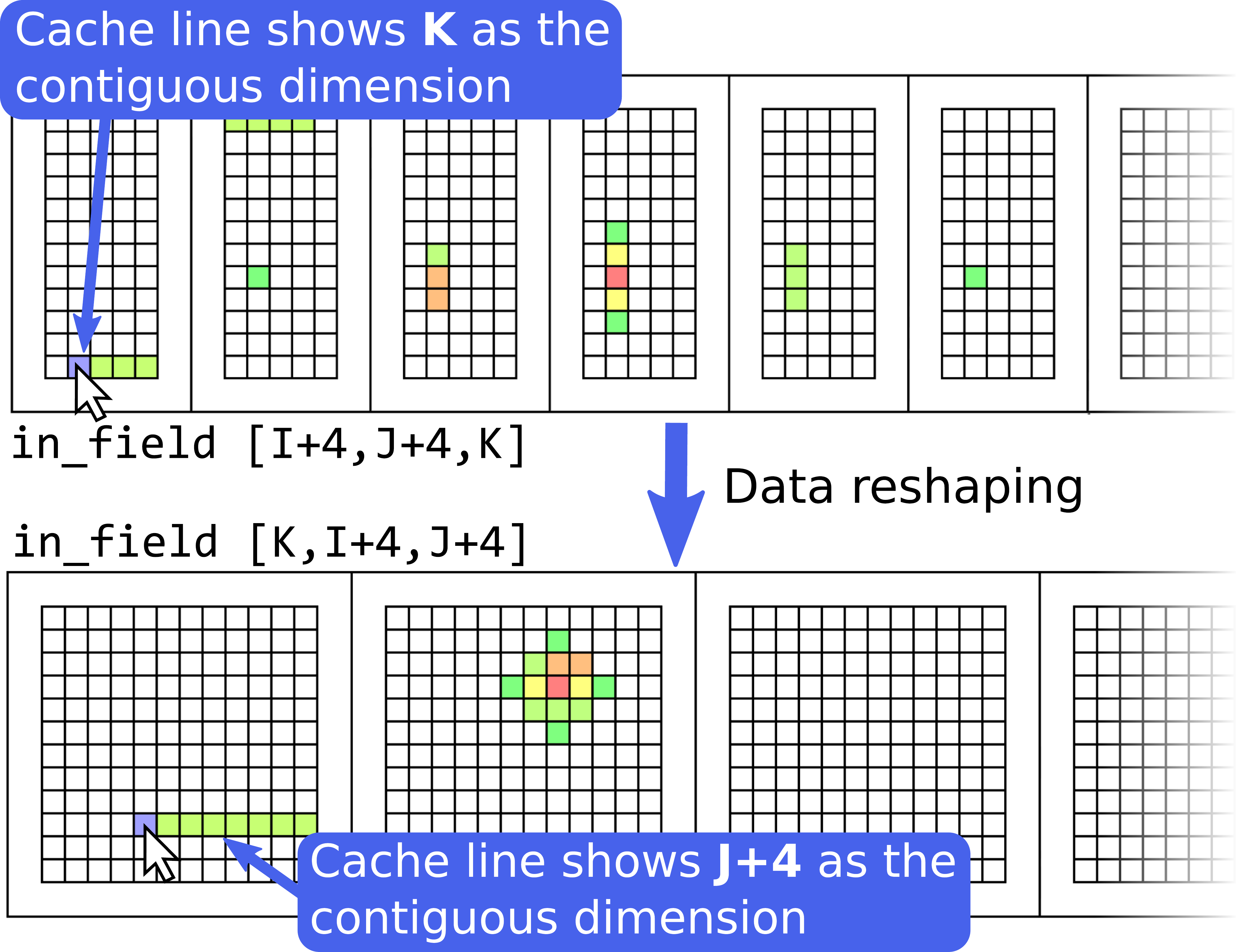}
		\caption{Accesses to $\mathtt{in\_field}$ for $i=1$, $j=5$, and $k=1$, before and after reshaping.}
		\label{fig:hdiff-bad-layout}
	\end{subfigure}\hfill
	\begin{subfigure}[b]{.32\linewidth}
		\centering
		\includegraphics[width=\linewidth]{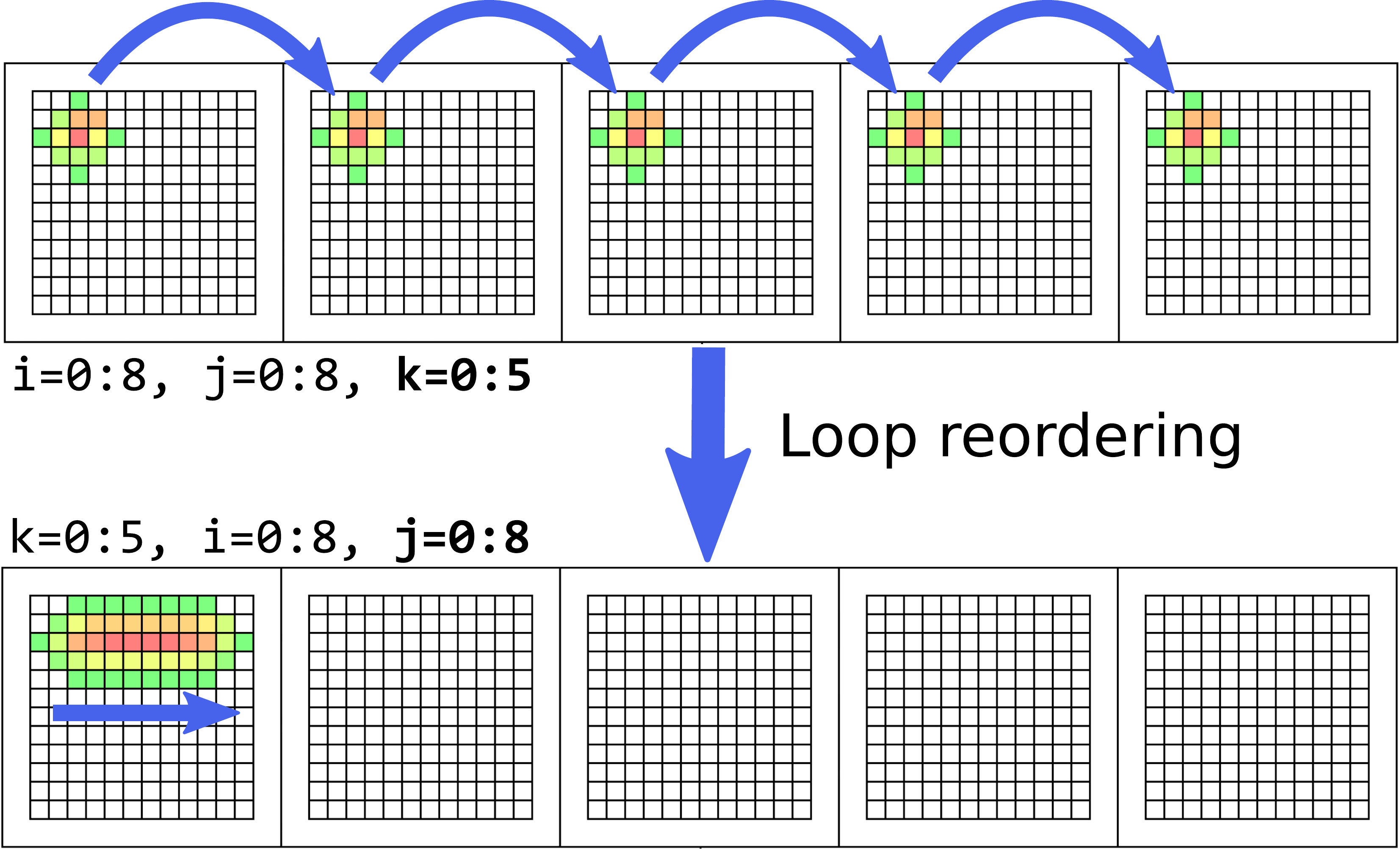}
		\caption{Access pattern on $\mathtt{in\_field}$ when iterating through the innermost loop, before and after reordering.}
		\label{fig:hdiff-loop-dim-shuffle}
	\end{subfigure}\hfill
	\begin{subfigure}[b]{.34\linewidth}
		\centering
		\includegraphics[width=\linewidth]{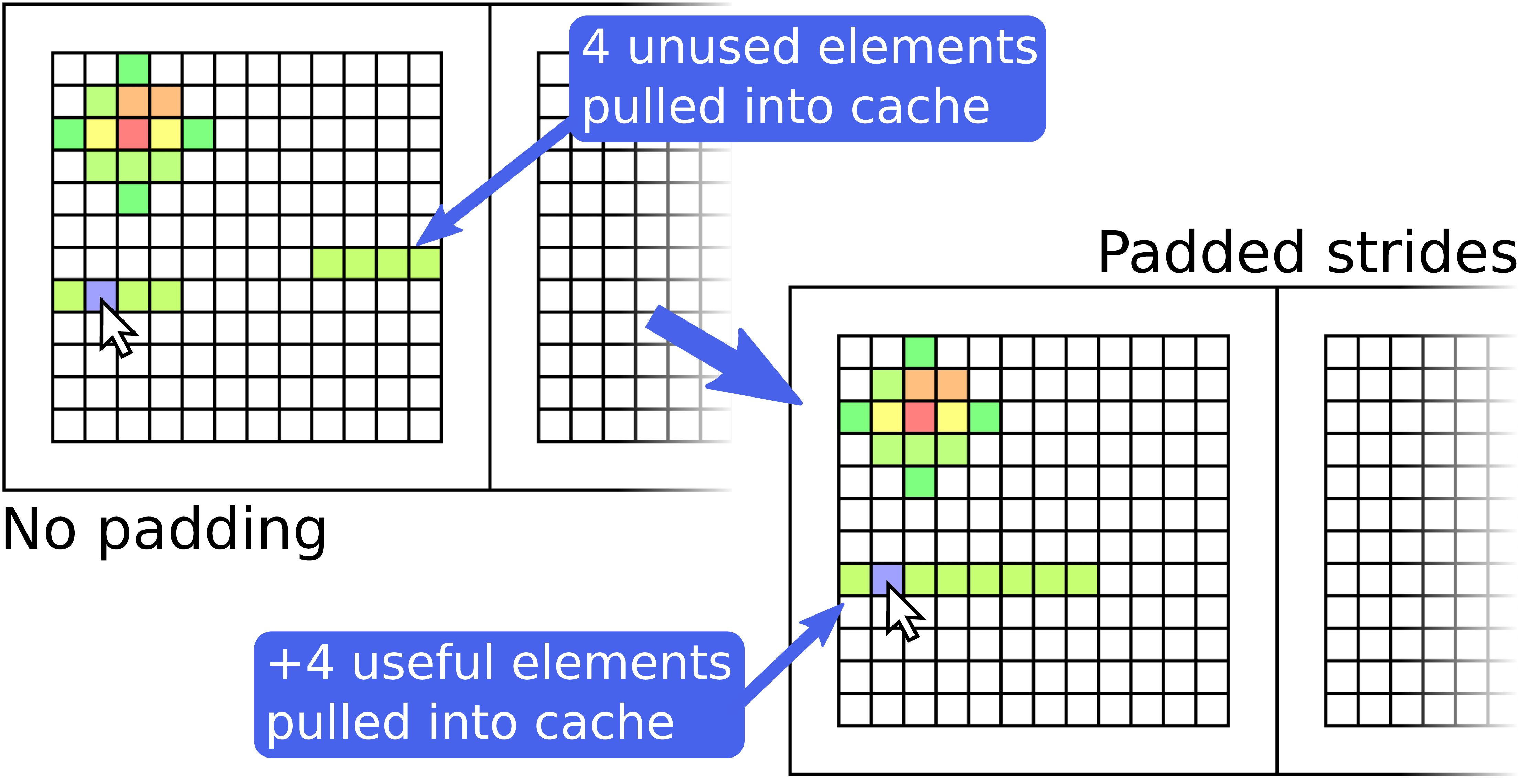}
		\caption{Introducing padding by increasing the strides along the second matrix dimension improves spatial locality.}
		\label{fig:hdiff-padding}
	\end{subfigure}
	\caption{Analysis of horizontal diffusion during individual steps of the tuning process.}
	\label{fig:hdiff-tuning}
\end{figure*}

Using only information directly exposed through our visualization, we get \textbf{speedups of $\mathbf{7.1 \times}$ to $\mathbf{30.2 \times}$}
depending on the target system, as shown in \autoref{tab:case-studies-results}.

\subsection{Horizontal Diffusion}\label{ss:hdiff-case-study}
To demonstrate the analysis capabilities of our visualization's close-up, local view, we walk through the tuning process of
horizontal diffusion (or \emph{hdiff}), which is a stencil composition that plays an important role in weather and climate models~\cite{Gysi2021}.
We take an implementation of this application from the high-performance NumPy~\cite{numpy} benchmarking suite NPBench~\cite{Ziogas2021}.
The program has three free parameters $I$, $J$, and $K$ and operates on two inputs $\mathtt{in\_field}\in \mathbb{R}^{I+4\!\times\!J+4\!\times\!K}$ and $\mathtt{coeff} \in \mathbb{R}^{I\!\times\!J\!\times\!K}$, and an output $\mathtt{out\_field}\in\mathbb{R}^{I\!\times\!J\!\times\!K}$.
We evaluate the application using the same parameter sizes used in the NPBench paper~\cite{Ziogas2021}, with $I=J=256$ and $K=160$, which represent a typical per-node scenario in a cluster running weather and climate simulations~\cite{Gysi2021}.
The default NumPy implementation in NPBench serves as our baseline.

To start the analysis process in the local view, we parameterize the program with smaller input parameters $I = J = 8$ and $K = 5$, which represents a $\frac{1}{32}$ scaled version of the full-sized program.
These parameters can be chosen arbitrarily, but scaling all input parameters down by a common factor ensures that all analysis parameters maintain the same size relative to each other.
The full resulting graphical representation of the program, which can be represented as one 3-dimensional loop operating on two input and one output data containers, can be seen in Fig.~\ref{fig:hdiff-progression} (left).
We further set the cache line size to 64 bytes to reflect our target architecture,
and since the program operates on double-precision floating point values, the value size is 8 bytes.

By moving any one of the loop sliders, or by hovering over or clicking on any output data element,
the corresponding access pattern on the input data containers is shown.
This access pattern shows that each loop iteration accesses a number of elements from the $\mathtt{in\_field}$ input, as shown at the top of Fig.~\ref{fig:hdiff-bad-layout}.
If these accesses occur far apart in memory, this leads to poor spatial reuse, since a large number of separate cache lines would have to be accessed for one operation.
Clicking any data element with the cache line overlay enabled exposes the data layout of the $\mathtt{in\_field}$ container.
The highlighted cache line shows that the container uses row-major ordering, which implies that dimension $J+4$, over which the accesses are spread out, is a non-contiguous dimension, and that the accesses are spread out in memory with poor spatial locality.
Reshaping $\mathtt{in\_field}$ from $[I+4, J+4, K]$ to $[K, I+4, J+4]$ visibly improves the access pattern, with all accesses now occurring much closer to each other in memory, as shown in the bottom half of Fig.~\ref{fig:hdiff-bad-layout}.
This optimization step further comes with a visible reduction in the number of cache misses, and consequently almost halves the amount of data being requested from main memory for $\mathtt{in\_field}$, as seen in Fig.~\ref{fig:hdiff-progression}.

While the accesses per loop iteration are now physically closer together, by playing back the access pattern animation or moving the sliders, the overlay reveals a new problem.
The innermost loop, $k\in[0,4]$, now iterates over a non-contiguous dimension, as shown in the top of Fig.~\ref{fig:hdiff-loop-dim-shuffle}.
A simple re-ordering of the loops, such that $k\in[0,4]$ becomes the outer-most loop, addresses this problem and improves the access pattern, as demonstrated in Fig.~\ref{fig:hdiff-loop-dim-shuffle} (bottom).
Fig.~\ref{fig:hdiff-progression} shows a further reduction of both cache misses and the number of data bytes moved following this optimization step.

Finally, the cache line visualization shown in Fig.~\ref{fig:hdiff-padding} (top) highlights that accesses to some of the first elements in individual rows are located on cache lines that wrap around from the previous row.
Knowing the access pattern of an individual loop iteration in horizontal diffusion (bottom of Fig.~\ref{fig:hdiff-bad-layout}), an iteration accessing these elements would simultaneously require accesses to more elements located in the same row.
The elements residing in the previous row are consequently unused and pollute the cache.
By increasing the strides along the second data dimension to a number divisible by the cache line size, we can introduce post-padding to align each individual data row to the cache line.
As shown in the bottom half of Fig.~\ref{fig:hdiff-padding}, this improves spatial locality by pulling elements into the cache that will be accessed in the same loop iteration.

We benchmark the optimized application and observe between a $51.2 \times$ and $151.4 \times$ improvement over the original NumPy implementation, depending on the target system, as shown in \autoref{tab:case-studies-results}.
This speedup through optimizations based on information obtained from our visualization, was attained \emph{without requiring profiling} or the analysis of hardware counters, and even \textbf{outperforms the fastest CPU-based result measured in NPBench~\cite{Ziogas2021} by between $\mathbf{5.7 \times}$ and $\mathbf{7.2 \times}$}, depending on the target system.

We provide three short supplementary videos demonstrating the analysis process for \textit{hdiff} when identifying the suboptimal memory layout\footnote{Video `Visualizing Data Layouts' (\url{https://youtu.be/H5DVE31-CW8})}, loop order\footnote{Video `Visualizing Data Access Patterns' (\url{https://youtu.be/cSlXTjqDxrk})}, and improper alignment\footnote{Video `Visualizing Spatial Locality' (\url{https://youtu.be/tZrpRt_6Yi4})}.

\section{Related Work}\label{s:related-work}
Several other works exist with the goal of assisting in the performance analysis workflow by either visualizing performance data, exploring data reuse with data-centric analysis, or obtaining performance insights through simulation.
Many approaches combine a subset of these techniques to provide visualization-augmented, data-centric analysis, but require lengthy program executions to obtain instrumentation data.
Even simulation- and modeling-based approaches often rely on memory traces obtained through program execution.
Existing approaches further often show performance metrics and observations separate from the program definition, requiring context switching between analysis and subsequent optimization steps.

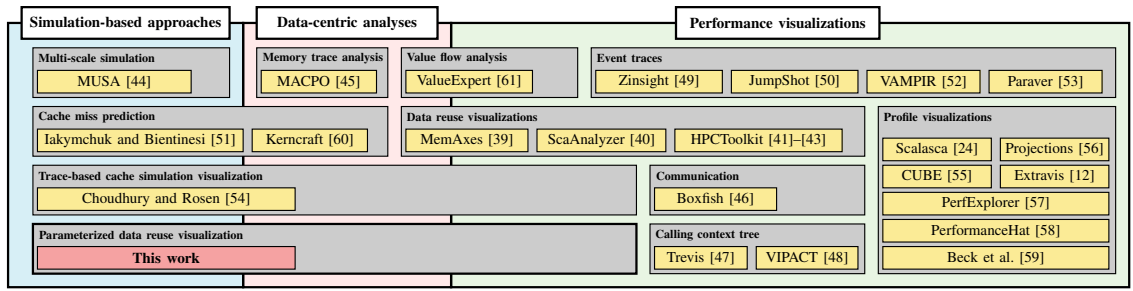
\begin{figure*}
	\centering
	\scalebox{0.57}{\input{fig/related_work_overview}}
	\caption{Related work overview.}
	\label{fig:related-work}
\end{figure*}

Our solution provides a holistic approach that attempts to remove the need for program executions entirely, enabling a more interactive performance optimization process.
We visualize memory access and reuse behavior on multiple scales, directly in-situ on an intermediate program representation that can be used for subsequent optimizations.
An overview of the related works can be seen in Fig.~\ref{fig:related-work}.

\paragraph{\textbf{Performance Visualization}}\label{p:perf-viz-related}
Many tools have been created to expose a program's performance characteristics by aggregating and visualizing profiling and tracing data.
Zinsight~\cite{DePauw2010} visualizes large event traces with a set of independent views that extract event statistics and patterns.
Event traces can also be explored using JumpShot~\cite{Zaki1997}, VAMPIR~\cite{Nagel1996}, or Paraver~\cite{Pillet1995}, which offer different approaches to navigating these large traces.
JumpShot, VAMPIR, and Paraver are all part of the larger TAU parallel performance analysis system~\cite{Shende2006a}, which further includes PerfExplorer~\cite{Huck2005}, a tool to summarize parallel profiles with a set of 2D and 3D graph visualizations.
In a similar manner, Projections~\cite{Kale2006} visualizes profile data in summary graphs, detailed timeline views, and other graphs, employing the help of overviews to facilitate easier navigation.
Projections further allows comparing profiles gathered from different runs side-by-side.
The use of overviews to help navigate large profiles can also be seen in Extravis~\cite{Cornelissen2007}, which shows call relations in a circular bundle view.
Scalasca~\cite{Geimer2010} and its visualization component CUBE~\cite{Geimer2008} have also been created with the goal of visualizing profiles obtained from programs using many thousands of processors.
PerformanceHat~\cite{Cito2019} and Beck et al.~\cite{Beck2013} visualize execution time measurements extracted from profiles directly in the code editor, using icons, color highlights, and tooltips in the source code.
Some visualizations, such as Trevis~\cite{Adamoli2010} and VIPACT~\cite{Nguyen2017}, focus on extracting and exploring full calling context trees from profile traces, while others, e.g., Boxfish~\cite{Isaacs2012}, extract communication behavior and visualize that on the network topology.

\paragraph{\textbf{Data-Centric Analysis}}\label{p:data-centric-related}
Solutions which focus specifically on data movement and reuse analysis include
MACPO~\cite{Rane2014}, which obtains memory traces and uses those to compute access behavior metrics for source-level data structures in C, C++, and FORTRAN, using cache models geared towards execution on multi-core chips.
ValueExpert~\cite{Zhou2022} constructs value flow graphs for GPU applications to help with detecting inefficient value-related patterns and guiding subsequent optimization decisions.
MemAxes~\cite{Gimenez2018} facilitates analysis by visualizing memory performance obtained from traces, and performing a scoring of obtained performance metrics to guide the user's attention.
ScaAnalyzer~\cite{Liu2015} and other augmentations~\cite{Liu2013, Liu2014} made to the HPCToolkit~\cite{Adhianto2010} architecture support data-centric profiling of parallel programs, and attribute and visualize the recorded metrics to pinpoint scaling losses and other performance bottlenecks due to memory access behaviors.

\paragraph{\textbf{Simulation-Based Approaches}}\label{p:simulation-related}
Other works based on performance simulation predict performance behavior to avoid runtime-based analyses, but they typically require memory traces obtained through execution for their simulations.
MUSA~\cite{Machines2016} offers a multi-scale simulation approach to analyze inter-node communication and intra-node microarchitecture interactions.
Iakymchuck and Bientinesi~\cite{Iakymchuk2012} have constructed a performance model that can accurately predict cache misses for fundamental linear algebra operations on Intel and AMD processors, and have managed to combine individual models to predict the performance of BLAS subroutines like matrix factorization.
Kerncraft~\cite{Hammer2017} follows a similar direction by constructing roof line and execution-cache-memory models for loop nests in stencil codes.
Choudhury and Rosen~\cite{Choudhury2011} created an animated visualization that shows memory transactions based on a memory reference trace, and uses a cache simulator to visualize data elements according to their simulated cache location.

\section{Discussion}\label{s:discussion}
The proposed visualization is designed to be flexible and extensible.
In the following, we discuss potential extensions, advantages, and limitations in the approach.

\paragraph{Cache Model}
Our tool performs a general-purpose cache miss estimation based on the simulated access patterns.
However, this estimation is separate from the visualization itself.
Cache miss predictions for specific architectures could be derived from the simulated access patterns using different, more hardware-specific back-ends, such as Kerncraft~\cite{Hammer2017}, while leveraging the same visual exploration and analysis methods demonstrated in our visualization.

\paragraph{Remote Analysis}
By using Visual Studio Code~\cite{vscode} as the basis for the implementation of our visualization tool, we can leverage the remote development feature in the code editor to perform analyses and optimizations directly on target machines, such as the compute or login nodes of supercomputers.
This further increases the interactivity of the optimization workflow, by providing intuitive visual analyses even for remote scenarios.

\paragraph{Program Parameterization}
Since data access patterns for regular programs do not depend on specific values of input parameters, finding good parameters to analyze a program in the local view is a matter of choosing what is easiest for the user to observe data access and reuse behavior.
A good strategy is to keep these parameters small, because this keeps simulation times short when transitioning to the local view, and makes analysis easier by limiting the visual search space and cognitive load required.
While the exact impact of individual optimizations may differ between the small-scale, parameterized setting and the full-scale scenario, our case study results demonstrate that the general insights obtained in the small-scale setting are helpful in uncovering key performance problems that transfer to the full-scale scenario.

This visualization approach could also be used to simulate and analyze the full-sized parameters.
However, this would significantly increase simulation times, and would require aggregating multiple data elements in one visual tile to avoid making the visualization harder to interpret.

\paragraph{Limitations}
A key limiting factor in our approach is in the analysis of dynamic or irregular programs.
For this class of programs, exact data movement and access information cannot usually be determined statically or through small-scale, parameterized simulations.
However, the global and local visualization techniques employed in our tool can similarly be used to analyze and explore traditional instrumentation data for such applications, allowing for a comparable, though less interactive optimization workflow.

\section{Conclusion}\label{s:conclusion}
We have implemented a performance visualization tool in Visual Studio Code that exposes critical performance characteristics to the user, which are generally not directly visible in traditional source code and require closer analysis, such as data layout and movement, and spatial or temporal locality.
By leveraging a combination of static dataflow analysis and small-scale simulations, our approach avoids costly profiling or instrumentation runs that slow down the tuning process.
In combination with visualizing results in-situ, directly on a graphical program representation, this facilitates a more interactive and streamlined optimization process.

Two case studies on the optimization of real-world applications demonstrate the effectiveness of our visualization in both global data movement reduction, and fine-grained data reuse optimizations.
Our tool demonstrates the ability to inform impactful tuning decisions on multiple levels, enabling optimization decisions that lead to a speedup of up to $7.2 \times$ compared to the state of the art on an HPC benchmark.
With data movement optimizations increasingly becoming a crucial part of performance tuning in HPC applications, we hope our approach to an interactive analysis process inspires further work on streamlined performance engineering processes.

\section*{Acknowledgments}
This project received funding from the European Research Council (ERC) \includegraphics[height=1em]{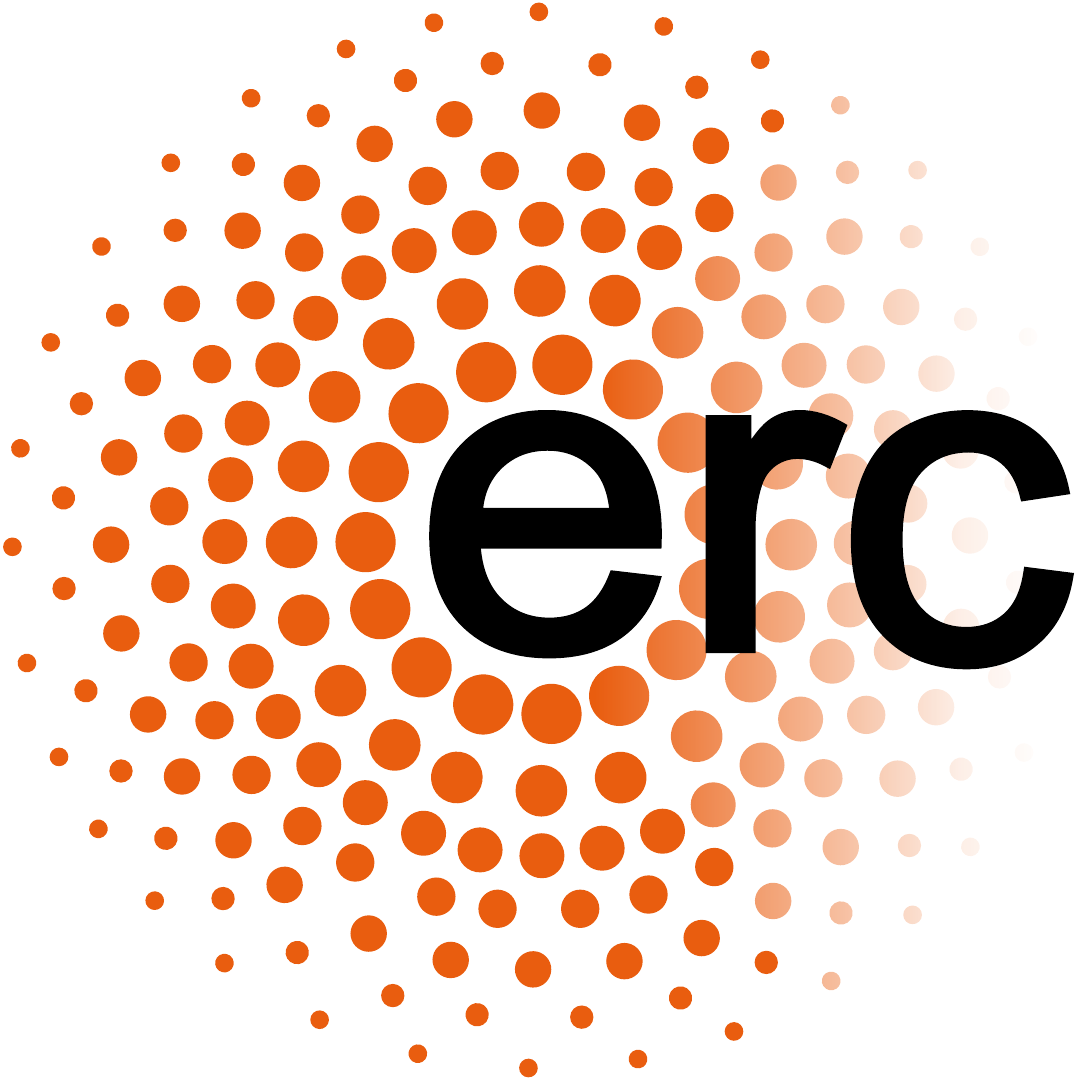}
grant PSAP, grant agreement No. 101002047, and the European Union's Horizon Europe programme
DEEP-SEA, grant agreement No. 955606.
P.S. and T.B.N. are supported by the Swiss National Science Foundation (Ambizione Project No. 185778).
The authors also wish to acknowledge the Swiss National Supercomputing Centre (CSCS) for access and support
of the computational resources.

\bibliographystyle{IEEEtran}
\input{paper.bbl}

\end{document}

%% file: fig/related_work_overview.tex
\ifx\du\undefined
  \newlength{\du}
\fi
\setlength{\du}{15\unitlength}
\begin{tikzpicture}
\pgftransformxscale{1.000000}
\pgftransformyscale{-1.000000}
\definecolor{dialinecolor}{rgb}{0.000000, 0.000000, 0.000000}
\pgfsetstrokecolor{dialinecolor}
\definecolor{dialinecolor}{rgb}{1.000000, 1.000000, 1.000000}
\pgfsetfillcolor{dialinecolor}
\pgfsetlinewidth{0.100000\du}
\pgfsetdash{}{0pt}
\pgfsetdash{}{0pt}
\pgfsetmiterjoin
\definecolor{dialinecolor}{rgb}{0.909804, 0.960784, 0.890196}
\pgfsetfillcolor{dialinecolor}
\fill (29.800000\du,1.700000\du)--(29.800000\du,13.400000\du)--(59.800000\du,13.400000\du)--(59.800000\du,1.700000\du)--cycle;
\definecolor{dialinecolor}{rgb}{0.000000, 0.000000, 0.000000}
\pgfsetstrokecolor{dialinecolor}
\draw (29.800000\du,1.700000\du)--(29.800000\du,13.400000\du)--(59.800000\du,13.400000\du)--(59.800000\du,1.700000\du)--cycle;
\pgfsetlinewidth{0.100000\du}
\pgfsetdash{}{0pt}
\pgfsetdash{}{0pt}
\pgfsetmiterjoin
\definecolor{dialinecolor}{rgb}{1.000000, 0.913725, 0.913725}
\pgfsetfillcolor{dialinecolor}
\fill (20.600000\du,1.700000\du)--(20.600000\du,13.400000\du)--(29.800000\du,13.400000\du)--(29.800000\du,1.700000\du)--cycle;
\definecolor{dialinecolor}{rgb}{0.000000, 0.000000, 0.000000}
\pgfsetstrokecolor{dialinecolor}
\draw (20.600000\du,1.700000\du)--(20.600000\du,13.400000\du)--(29.800000\du,13.400000\du)--(29.800000\du,1.700000\du)--cycle;
\pgfsetlinewidth{0.100000\du}
\pgfsetdash{}{0pt}
\pgfsetdash{}{0pt}
\pgfsetmiterjoin
\definecolor{dialinecolor}{rgb}{0.843137, 0.937255, 0.964706}
\pgfsetfillcolor{dialinecolor}
\fill (10.200000\du,1.700000\du)--(10.200000\du,13.400000\du)--(20.600000\du,13.400000\du)--(20.600000\du,1.700000\du)--cycle;
\definecolor{dialinecolor}{rgb}{0.000000, 0.000000, 0.000000}
\pgfsetstrokecolor{dialinecolor}
\draw (10.200000\du,1.700000\du)--(10.200000\du,13.400000\du)--(20.600000\du,13.400000\du)--(20.600000\du,1.700000\du)--cycle;
\pgfsetlinewidth{0.050000\du}
\pgfsetdash{}{0pt}
\pgfsetdash{}{0pt}
\pgfsetmiterjoin
\definecolor{dialinecolor}{rgb}{0.800000, 0.800000, 0.800000}
\pgfsetfillcolor{dialinecolor}
\fill (38.600000\du,8.000000\du)--(38.600000\du,10.200000\du)--(48.100000\du,10.200000\du)--(48.100000\du,8.000000\du)--cycle;
\definecolor{dialinecolor}{rgb}{0.000000, 0.000000, 0.000000}
\pgfsetstrokecolor{dialinecolor}
\draw (38.600000\du,8.000000\du)--(38.600000\du,10.200000\du)--(48.100000\du,10.200000\du)--(48.100000\du,8.000000\du)--cycle;
\pgfsetlinewidth{0.050000\du}
\pgfsetdash{}{0pt}
\pgfsetdash{}{0pt}
\pgfsetmiterjoin
\definecolor{dialinecolor}{rgb}{0.800000, 0.800000, 0.800000}
\pgfsetfillcolor{dialinecolor}
\fill (21.200000\du,2.800000\du)--(21.200000\du,5.000000\du)--(27.000000\du,5.000000\du)--(27.000000\du,2.800000\du)--cycle;
\definecolor{dialinecolor}{rgb}{0.000000, 0.000000, 0.000000}
\pgfsetstrokecolor{dialinecolor}
\draw (21.200000\du,2.800000\du)--(21.200000\du,5.000000\du)--(27.000000\du,5.000000\du)--(27.000000\du,2.800000\du)--cycle;
\definecolor{dialinecolor}{rgb}{0.000000, 0.000000, 0.000000}
\pgfsetstrokecolor{dialinecolor}
\node[anchor=west] at (13.400000\du,4.900000\du){};
\definecolor{dialinecolor}{rgb}{0.000000, 0.000000, 0.000000}
\pgfsetstrokecolor{dialinecolor}
\node[anchor=west] at (26.500000\du,3.000000\du){};
\pgfsetlinewidth{0.100000\du}
\pgfsetdash{}{0pt}
\pgfsetdash{}{0pt}
\pgfsetmiterjoin
\definecolor{dialinecolor}{rgb}{1.000000, 1.000000, 1.000000}
\pgfsetfillcolor{dialinecolor}
\fill (21.200000\du,1.000000\du)--(21.200000\du,2.400000\du)--(29.200000\du,2.400000\du)--(29.200000\du,1.000000\du)--cycle;
\definecolor{dialinecolor}{rgb}{0.000000, 0.000000, 0.000000}
\pgfsetstrokecolor{dialinecolor}
\draw (21.200000\du,1.000000\du)--(21.200000\du,2.400000\du)--(29.200000\du,2.400000\du)--(29.200000\du,1.000000\du)--cycle;
\definecolor{dialinecolor}{rgb}{0.000000, 0.000000, 0.000000}
\pgfsetstrokecolor{dialinecolor}
\node at (25.200000\du,1.700000\du){\textbf{Data-centric analyses}};
\pgfsetlinewidth{0.100000\du}
\pgfsetdash{}{0pt}
\pgfsetdash{}{0pt}
\pgfsetmiterjoin
\definecolor{dialinecolor}{rgb}{1.000000, 1.000000, 1.000000}
\pgfsetfillcolor{dialinecolor}
\fill (39.749900\du,1.000000\du)--(39.749900\du,2.400000\du)--(48.749900\du,2.400000\du)--(48.749900\du,1.000000\du)--cycle;
\definecolor{dialinecolor}{rgb}{0.000000, 0.000000, 0.000000}
\pgfsetstrokecolor{dialinecolor}
\draw (39.749900\du,1.000000\du)--(39.749900\du,2.400000\du)--(48.749900\du,2.400000\du)--(48.749900\du,1.000000\du)--cycle;
\definecolor{dialinecolor}{rgb}{0.000000, 0.000000, 0.000000}
\pgfsetstrokecolor{dialinecolor}
\node at (44.249900\du,1.700000\du){\textbf{Performance visualizations}};
\definecolor{dialinecolor}{rgb}{0.000000, 0.000000, 0.000000}
\pgfsetstrokecolor{dialinecolor}
\node[anchor=west] at (41.800000\du,5.600000\du){};
\definecolor{dialinecolor}{rgb}{0.000000, 0.000000, 0.000000}
\pgfsetstrokecolor{dialinecolor}
\node[anchor=west] at (37.171400\du,3.000000\du){};
\definecolor{dialinecolor}{rgb}{0.000000, 0.000000, 0.000000}
\pgfsetstrokecolor{dialinecolor}
\node[anchor=west] at (38.600000\du,8.470000\du){\footnotesize{\textbf{Communication}}};
\definecolor{dialinecolor}{rgb}{0.000000, 0.000000, 0.000000}
\pgfsetstrokecolor{dialinecolor}
\node[anchor=west] at (21.200000\du,3.270000\du){\footnotesize{\textbf{Memory trace analysis}}};
\pgfsetlinewidth{0.100000\du}
\pgfsetdash{}{0pt}
\pgfsetdash{}{0pt}
\pgfsetmiterjoin
\definecolor{dialinecolor}{rgb}{1.000000, 1.000000, 1.000000}
\pgfsetfillcolor{dialinecolor}
\fill (10.800000\du,1.000000\du)--(10.800000\du,2.400000\du)--(20.000000\du,2.400000\du)--(20.000000\du,1.000000\du)--cycle;
\definecolor{dialinecolor}{rgb}{0.000000, 0.000000, 0.000000}
\pgfsetstrokecolor{dialinecolor}
\draw (10.800000\du,1.000000\du)--(10.800000\du,2.400000\du)--(20.000000\du,2.400000\du)--(20.000000\du,1.000000\du)--cycle;
\definecolor{dialinecolor}{rgb}{0.000000, 0.000000, 0.000000}
\pgfsetstrokecolor{dialinecolor}
\node[anchor=west] at (15.400000\du,1.700000\du){};
\definecolor{dialinecolor}{rgb}{0.000000, 0.000000, 0.000000}
\pgfsetstrokecolor{dialinecolor}
\node[anchor=west] at (15.400000\du,1.700000\du){};
\definecolor{dialinecolor}{rgb}{0.000000, 0.000000, 0.000000}
\pgfsetstrokecolor{dialinecolor}
\node at (15.400000\du,1.700000\du){\textbf{Simulation-based approaches}};
\pgfsetlinewidth{0.050000\du}
\pgfsetdash{}{0pt}
\pgfsetdash{}{0pt}
\pgfsetmiterjoin
\definecolor{dialinecolor}{rgb}{0.800000, 0.800000, 0.800000}
\pgfsetfillcolor{dialinecolor}
\fill (27.600000\du,5.400000\du)--(27.600000\du,7.600000\du)--(48.100000\du,7.600000\du)--(48.100000\du,5.400000\du)--cycle;
\definecolor{dialinecolor}{rgb}{0.000000, 0.000000, 0.000000}
\pgfsetstrokecolor{dialinecolor}
\draw (27.600000\du,5.400000\du)--(27.600000\du,7.600000\du)--(48.100000\du,7.600000\du)--(48.100000\du,5.400000\du)--cycle;
\definecolor{dialinecolor}{rgb}{0.000000, 0.000000, 0.000000}
\pgfsetstrokecolor{dialinecolor}
\node[anchor=west] at (27.600000\du,5.870000\du){\footnotesize{\textbf{Data reuse visualizations}}};
\pgfsetlinewidth{0.050000\du}
\pgfsetdash{}{0pt}
\pgfsetdash{}{0pt}
\pgfsetmiterjoin
\definecolor{dialinecolor}{rgb}{0.984314, 0.921569, 0.588235}
\pgfsetfillcolor{dialinecolor}
\fill (27.800000\du,6.400000\du)--(27.800000\du,7.400000\du)--(33.200000\du,7.400000\du)--(33.200000\du,6.400000\du)--cycle;
\definecolor{dialinecolor}{rgb}{0.000000, 0.000000, 0.000000}
\pgfsetstrokecolor{dialinecolor}
\draw (27.800000\du,6.400000\du)--(27.800000\du,7.400000\du)--(33.200000\du,7.400000\du)--(33.200000\du,6.400000\du)--cycle;
\definecolor{dialinecolor}{rgb}{0.000000, 0.000000, 0.000000}
\pgfsetstrokecolor{dialinecolor}
\node at (30.500000\du,6.900000\du){MemAxes \cite{Gimenez2018}};
\pgfsetlinewidth{0.050000\du}
\pgfsetdash{}{0pt}
\pgfsetdash{}{0pt}
\pgfsetmiterjoin
\definecolor{dialinecolor}{rgb}{0.984314, 0.921569, 0.588235}
\pgfsetfillcolor{dialinecolor}
\fill (33.600000\du,6.400000\du)--(33.600000\du,7.400000\du)--(39.300000\du,7.400000\du)--(39.300000\du,6.400000\du)--cycle;
\definecolor{dialinecolor}{rgb}{0.000000, 0.000000, 0.000000}
\pgfsetstrokecolor{dialinecolor}
\draw (33.600000\du,6.400000\du)--(33.600000\du,7.400000\du)--(39.300000\du,7.400000\du)--(39.300000\du,6.400000\du)--cycle;
\definecolor{dialinecolor}{rgb}{0.000000, 0.000000, 0.000000}
\pgfsetstrokecolor{dialinecolor}
\node at (36.450000\du,6.900000\du){ScaAnalyzer \cite{Liu2015}};
\pgfsetlinewidth{0.050000\du}
\pgfsetdash{}{0pt}
\pgfsetdash{}{0pt}
\pgfsetmiterjoin
\definecolor{dialinecolor}{rgb}{0.984314, 0.921569, 0.588235}
\pgfsetfillcolor{dialinecolor}
\fill (39.700000\du,6.400000\du)--(39.700000\du,7.400000\du)--(47.100000\du,7.400000\du)--(47.100000\du,6.400000\du)--cycle;
\definecolor{dialinecolor}{rgb}{0.000000, 0.000000, 0.000000}
\pgfsetstrokecolor{dialinecolor}
\draw (39.700000\du,6.400000\du)--(39.700000\du,7.400000\du)--(47.100000\du,7.400000\du)--(47.100000\du,6.400000\du)--cycle;
\definecolor{dialinecolor}{rgb}{0.000000, 0.000000, 0.000000}
\pgfsetstrokecolor{dialinecolor}
\node at (43.400000\du,6.900000\du){HPCToolkit \cite{Liu2013, Liu2014, Adhianto2010}};
\pgfsetlinewidth{0.050000\du}
\pgfsetdash{}{0pt}
\pgfsetdash{}{0pt}
\pgfsetmiterjoin
\definecolor{dialinecolor}{rgb}{0.800000, 0.800000, 0.800000}
\pgfsetfillcolor{dialinecolor}
\fill (11.300000\du,2.800000\du)--(11.300000\du,5.000000\du)--(20.000000\du,5.000000\du)--(20.000000\du,2.800000\du)--cycle;
\definecolor{dialinecolor}{rgb}{0.000000, 0.000000, 0.000000}
\pgfsetstrokecolor{dialinecolor}
\draw (11.300000\du,2.800000\du)--(11.300000\du,5.000000\du)--(20.000000\du,5.000000\du)--(20.000000\du,2.800000\du)--cycle;
\definecolor{dialinecolor}{rgb}{0.000000, 0.000000, 0.000000}
\pgfsetstrokecolor{dialinecolor}
\node[anchor=west] at (11.300000\du,3.270000\du){\footnotesize{\textbf{Multi-scale simulation}}};
\pgfsetlinewidth{0.050000\du}
\pgfsetdash{}{0pt}
\pgfsetdash{}{0pt}
\pgfsetmiterjoin
\definecolor{dialinecolor}{rgb}{0.984314, 0.921569, 0.588235}
\pgfsetfillcolor{dialinecolor}
\fill (11.500000\du,3.800000\du)--(11.500000\du,4.800000\du)--(18.300000\du,4.800000\du)--(18.300000\du,3.800000\du)--cycle;
\definecolor{dialinecolor}{rgb}{0.000000, 0.000000, 0.000000}
\pgfsetstrokecolor{dialinecolor}
\draw (11.500000\du,3.800000\du)--(11.500000\du,4.800000\du)--(18.300000\du,4.800000\du)--(18.300000\du,3.800000\du)--cycle;
\definecolor{dialinecolor}{rgb}{0.000000, 0.000000, 0.000000}
\pgfsetstrokecolor{dialinecolor}
\node at (14.900000\du,4.300000\du){MUSA \cite{Machines2016}};
\pgfsetlinewidth{0.050000\du}
\pgfsetdash{}{0pt}
\pgfsetdash{}{0pt}
\pgfsetmiterjoin
\definecolor{dialinecolor}{rgb}{0.984314, 0.921569, 0.588235}
\pgfsetfillcolor{dialinecolor}
\fill (21.400000\du,3.800000\du)--(21.400000\du,4.800000\du)--(26.500000\du,4.800000\du)--(26.500000\du,3.800000\du)--cycle;
\definecolor{dialinecolor}{rgb}{0.000000, 0.000000, 0.000000}
\pgfsetstrokecolor{dialinecolor}
\draw (21.400000\du,3.800000\du)--(21.400000\du,4.800000\du)--(26.500000\du,4.800000\du)--(26.500000\du,3.800000\du)--cycle;
\definecolor{dialinecolor}{rgb}{0.000000, 0.000000, 0.000000}
\pgfsetstrokecolor{dialinecolor}
\node at (23.950000\du,4.300000\du){MACPO \cite{Rane2014}};
\pgfsetlinewidth{0.050000\du}
\pgfsetdash{}{0pt}
\pgfsetdash{}{0pt}
\pgfsetmiterjoin
\definecolor{dialinecolor}{rgb}{0.984314, 0.921569, 0.588235}
\pgfsetfillcolor{dialinecolor}
\fill (38.800000\du,9.000000\du)--(38.800000\du,10.000000\du)--(44.200000\du,10.000000\du)--(44.200000\du,9.000000\du)--cycle;
\definecolor{dialinecolor}{rgb}{0.000000, 0.000000, 0.000000}
\pgfsetstrokecolor{dialinecolor}
\draw (38.800000\du,9.000000\du)--(38.800000\du,10.000000\du)--(44.200000\du,10.000000\du)--(44.200000\du,9.000000\du)--cycle;
\definecolor{dialinecolor}{rgb}{0.000000, 0.000000, 0.000000}
\pgfsetstrokecolor{dialinecolor}
\node at (41.500000\du,9.500000\du){Boxfish \cite{Isaacs2012}};
\pgfsetlinewidth{0.050000\du}
\pgfsetdash{}{0pt}
\pgfsetdash{}{0pt}
\pgfsetmiterjoin
\definecolor{dialinecolor}{rgb}{0.800000, 0.800000, 0.800000}
\pgfsetfillcolor{dialinecolor}
\fill (38.600000\du,10.600000\du)--(38.600000\du,12.800000\du)--(48.100000\du,12.800000\du)--(48.100000\du,10.600000\du)--cycle;
\definecolor{dialinecolor}{rgb}{0.000000, 0.000000, 0.000000}
\pgfsetstrokecolor{dialinecolor}
\draw (38.600000\du,10.600000\du)--(38.600000\du,12.800000\du)--(48.100000\du,12.800000\du)--(48.100000\du,10.600000\du)--cycle;
\definecolor{dialinecolor}{rgb}{0.000000, 0.000000, 0.000000}
\pgfsetstrokecolor{dialinecolor}
\node[anchor=west] at (38.600000\du,11.070000\du){\footnotesize{\textbf{Calling context tree}}};
\pgfsetlinewidth{0.050000\du}
\pgfsetdash{}{0pt}
\pgfsetdash{}{0pt}
\pgfsetmiterjoin
\definecolor{dialinecolor}{rgb}{0.984314, 0.921569, 0.588235}
\pgfsetfillcolor{dialinecolor}
\fill (38.800000\du,11.600000\du)--(38.800000\du,12.600000\du)--(42.900000\du,12.600000\du)--(42.900000\du,11.600000\du)--cycle;
\definecolor{dialinecolor}{rgb}{0.000000, 0.000000, 0.000000}
\pgfsetstrokecolor{dialinecolor}
\draw (38.800000\du,11.600000\du)--(38.800000\du,12.600000\du)--(42.900000\du,12.600000\du)--(42.900000\du,11.600000\du)--cycle;
\definecolor{dialinecolor}{rgb}{0.000000, 0.000000, 0.000000}
\pgfsetstrokecolor{dialinecolor}
\node at (40.850000\du,12.100000\du){Trevis \cite{Adamoli2010}};
\pgfsetlinewidth{0.050000\du}
\pgfsetdash{}{0pt}
\pgfsetdash{}{0pt}
\pgfsetmiterjoin
\definecolor{dialinecolor}{rgb}{0.984314, 0.921569, 0.588235}
\pgfsetfillcolor{dialinecolor}
\fill (43.300000\du,11.600000\du)--(43.300000\du,12.600000\du)--(47.800000\du,12.600000\du)--(47.800000\du,11.600000\du)--cycle;
\definecolor{dialinecolor}{rgb}{0.000000, 0.000000, 0.000000}
\pgfsetstrokecolor{dialinecolor}
\draw (43.300000\du,11.600000\du)--(43.300000\du,12.600000\du)--(47.800000\du,12.600000\du)--(47.800000\du,11.600000\du)--cycle;
\definecolor{dialinecolor}{rgb}{0.000000, 0.000000, 0.000000}
\pgfsetstrokecolor{dialinecolor}
\node at (45.550000\du,12.100000\du){VIPACT \cite{Nguyen2017}};
\pgfsetlinewidth{0.050000\du}
\pgfsetdash{}{0pt}
\pgfsetdash{}{0pt}
\pgfsetmiterjoin
\definecolor{dialinecolor}{rgb}{0.800000, 0.800000, 0.800000}
\pgfsetfillcolor{dialinecolor}
\fill (36.000000\du,2.800000\du)--(36.000000\du,5.000000\du)--(59.200000\du,5.000000\du)--(59.200000\du,2.800000\du)--cycle;
\definecolor{dialinecolor}{rgb}{0.000000, 0.000000, 0.000000}
\pgfsetstrokecolor{dialinecolor}
\draw (36.000000\du,2.800000\du)--(36.000000\du,5.000000\du)--(59.200000\du,5.000000\du)--(59.200000\du,2.800000\du)--cycle;
\definecolor{dialinecolor}{rgb}{0.000000, 0.000000, 0.000000}
\pgfsetstrokecolor{dialinecolor}
\node[anchor=west] at (36.000000\du,3.270000\du){\footnotesize{\textbf{Event traces}}};
\pgfsetlinewidth{0.050000\du}
\pgfsetdash{}{0pt}
\pgfsetdash{}{0pt}
\pgfsetmiterjoin
\definecolor{dialinecolor}{rgb}{0.984314, 0.921569, 0.588235}
\pgfsetfillcolor{dialinecolor}
\fill (36.200000\du,3.800000\du)--(36.200000\du,4.800000\du)--(41.800000\du,4.800000\du)--(41.800000\du,3.800000\du)--cycle;
\definecolor{dialinecolor}{rgb}{0.000000, 0.000000, 0.000000}
\pgfsetstrokecolor{dialinecolor}
\draw (36.200000\du,3.800000\du)--(36.200000\du,4.800000\du)--(41.800000\du,4.800000\du)--(41.800000\du,3.800000\du)--cycle;
\definecolor{dialinecolor}{rgb}{0.000000, 0.000000, 0.000000}
\pgfsetstrokecolor{dialinecolor}
\node at (39.000000\du,4.300000\du){Zinsight \cite{DePauw2010}};
\pgfsetlinewidth{0.050000\du}
\pgfsetdash{}{0pt}
\pgfsetdash{}{0pt}
\pgfsetmiterjoin
\definecolor{dialinecolor}{rgb}{0.984314, 0.921569, 0.588235}
\pgfsetfillcolor{dialinecolor}
\fill (42.200000\du,3.800000\du)--(42.200000\du,4.800000\du)--(47.800000\du,4.800000\du)--(47.800000\du,3.800000\du)--cycle;
\definecolor{dialinecolor}{rgb}{0.000000, 0.000000, 0.000000}
\pgfsetstrokecolor{dialinecolor}
\draw (42.200000\du,3.800000\du)--(42.200000\du,4.800000\du)--(47.800000\du,4.800000\du)--(47.800000\du,3.800000\du)--cycle;
\definecolor{dialinecolor}{rgb}{0.000000, 0.000000, 0.000000}
\pgfsetstrokecolor{dialinecolor}
\node at (45.000000\du,4.300000\du){JumpShot \cite{Zaki1997}};
\pgfsetlinewidth{0.050000\du}
\pgfsetdash{}{0pt}
\pgfsetdash{}{0pt}
\pgfsetmiterjoin
\definecolor{dialinecolor}{rgb}{0.800000, 0.800000, 0.800000}
\pgfsetfillcolor{dialinecolor}
\fill (11.300000\du,5.400000\du)--(11.300000\du,7.600000\du)--(27.000000\du,7.600000\du)--(27.000000\du,5.400000\du)--cycle;
\definecolor{dialinecolor}{rgb}{0.000000, 0.000000, 0.000000}
\pgfsetstrokecolor{dialinecolor}
\draw (11.300000\du,5.400000\du)--(11.300000\du,7.600000\du)--(27.000000\du,7.600000\du)--(27.000000\du,5.400000\du)--cycle;
\definecolor{dialinecolor}{rgb}{0.000000, 0.000000, 0.000000}
\pgfsetstrokecolor{dialinecolor}
\node[anchor=west] at (11.300000\du,5.870000\du){\footnotesize{\textbf{Cache miss prediction}}};
\pgfsetlinewidth{0.050000\du}
\pgfsetdash{}{0pt}
\pgfsetdash{}{0pt}
\pgfsetmiterjoin
\definecolor{dialinecolor}{rgb}{0.984314, 0.921569, 0.588235}
\pgfsetfillcolor{dialinecolor}
\fill (11.500000\du,6.400000\du)--(11.500000\du,7.400000\du)--(20.600000\du,7.400000\du)--(20.600000\du,6.400000\du)--cycle;
\definecolor{dialinecolor}{rgb}{0.000000, 0.000000, 0.000000}
\pgfsetstrokecolor{dialinecolor}
\draw (11.500000\du,6.400000\du)--(11.500000\du,7.400000\du)--(20.600000\du,7.400000\du)--(20.600000\du,6.400000\du)--cycle;
\definecolor{dialinecolor}{rgb}{0.000000, 0.000000, 0.000000}
\pgfsetstrokecolor{dialinecolor}
\node at (16.050000\du,6.900000\du){Iakymchuk and Bientinesi \cite{Iakymchuk2012}};
\pgfsetlinewidth{0.050000\du}
\pgfsetdash{}{0pt}
\pgfsetdash{}{0pt}
\pgfsetmiterjoin
\definecolor{dialinecolor}{rgb}{0.984314, 0.921569, 0.588235}
\pgfsetfillcolor{dialinecolor}
\fill (48.200000\du,3.800000\du)--(48.200000\du,4.800000\du)--(53.200000\du,4.800000\du)--(53.200000\du,3.800000\du)--cycle;
\definecolor{dialinecolor}{rgb}{0.000000, 0.000000, 0.000000}
\pgfsetstrokecolor{dialinecolor}
\draw (48.200000\du,3.800000\du)--(48.200000\du,4.800000\du)--(53.200000\du,4.800000\du)--(53.200000\du,3.800000\du)--cycle;
\definecolor{dialinecolor}{rgb}{0.000000, 0.000000, 0.000000}
\pgfsetstrokecolor{dialinecolor}
\node at (50.700000\du,4.300000\du){VAMPIR \cite{Nagel1996}};
\pgfsetlinewidth{0.050000\du}
\pgfsetdash{}{0pt}
\pgfsetdash{}{0pt}
\pgfsetmiterjoin
\definecolor{dialinecolor}{rgb}{0.984314, 0.921569, 0.588235}
\pgfsetfillcolor{dialinecolor}
\fill (53.600000\du,3.800000\du)--(53.600000\du,4.800000\du)--(58.600000\du,4.800000\du)--(58.600000\du,3.800000\du)--cycle;
\definecolor{dialinecolor}{rgb}{0.000000, 0.000000, 0.000000}
\pgfsetstrokecolor{dialinecolor}
\draw (53.600000\du,3.800000\du)--(53.600000\du,4.800000\du)--(58.600000\du,4.800000\du)--(58.600000\du,3.800000\du)--cycle;
\definecolor{dialinecolor}{rgb}{0.000000, 0.000000, 0.000000}
\pgfsetstrokecolor{dialinecolor}
\node at (56.100000\du,4.300000\du){Paraver \cite{Pillet1995}};
\pgfsetlinewidth{0.050000\du}
\pgfsetdash{}{0pt}
\pgfsetdash{}{0pt}
\pgfsetmiterjoin
\definecolor{dialinecolor}{rgb}{0.800000, 0.800000, 0.800000}
\pgfsetfillcolor{dialinecolor}
\fill (11.300000\du,8.000000\du)--(11.300000\du,10.200000\du)--(38.000000\du,10.200000\du)--(38.000000\du,8.000000\du)--cycle;
\definecolor{dialinecolor}{rgb}{0.000000, 0.000000, 0.000000}
\pgfsetstrokecolor{dialinecolor}
\draw (11.300000\du,8.000000\du)--(11.300000\du,10.200000\du)--(38.000000\du,10.200000\du)--(38.000000\du,8.000000\du)--cycle;
\definecolor{dialinecolor}{rgb}{0.000000, 0.000000, 0.000000}
\pgfsetstrokecolor{dialinecolor}
\node[anchor=west] at (11.300000\du,8.470000\du){\footnotesize{\textbf{Trace-based cache simulation visualization}}};
\pgfsetlinewidth{0.050000\du}
\pgfsetdash{}{0pt}
\pgfsetdash{}{0pt}
\pgfsetmiterjoin
\definecolor{dialinecolor}{rgb}{0.984314, 0.921569, 0.588235}
\pgfsetfillcolor{dialinecolor}
\fill (11.500000\du,9.000000\du)--(11.500000\du,10.000000\du)--(22.900000\du,10.000000\du)--(22.900000\du,9.000000\du)--cycle;
\definecolor{dialinecolor}{rgb}{0.000000, 0.000000, 0.000000}
\pgfsetstrokecolor{dialinecolor}
\draw (11.500000\du,9.000000\du)--(11.500000\du,10.000000\du)--(22.900000\du,10.000000\du)--(22.900000\du,9.000000\du)--cycle;
\definecolor{dialinecolor}{rgb}{0.000000, 0.000000, 0.000000}
\pgfsetstrokecolor{dialinecolor}
\node at (17.200000\du,9.500000\du){Choudhury and Rosen \cite{Choudhury2011}};
\pgfsetlinewidth{0.100000\du}
\pgfsetdash{}{0pt}
\pgfsetdash{}{0pt}
\pgfsetmiterjoin
\definecolor{dialinecolor}{rgb}{0.800000, 0.800000, 0.800000}
\pgfsetfillcolor{dialinecolor}
\fill (11.300000\du,10.600000\du)--(11.300000\du,12.800000\du)--(38.000000\du,12.800000\du)--(38.000000\du,10.600000\du)--cycle;
\definecolor{dialinecolor}{rgb}{0.000000, 0.000000, 0.000000}
\pgfsetstrokecolor{dialinecolor}
\draw (11.300000\du,10.600000\du)--(11.300000\du,12.800000\du)--(38.000000\du,12.800000\du)--(38.000000\du,10.600000\du)--cycle;
\definecolor{dialinecolor}{rgb}{0.000000, 0.000000, 0.000000}
\pgfsetstrokecolor{dialinecolor}
\node[anchor=west] at (11.300000\du,11.070000\du){\footnotesize{\textbf{Parameterized data reuse visualization}}};
\pgfsetlinewidth{0.050000\du}
\pgfsetdash{}{0pt}
\pgfsetdash{}{0pt}
\pgfsetmiterjoin
\definecolor{dialinecolor}{rgb}{0.800000, 0.800000, 0.800000}
\pgfsetfillcolor{dialinecolor}
\fill (48.700000\du,5.400000\du)--(48.700000\du,12.800000\du)--(59.200000\du,12.800000\du)--(59.200000\du,5.400000\du)--cycle;
\definecolor{dialinecolor}{rgb}{0.000000, 0.000000, 0.000000}
\pgfsetstrokecolor{dialinecolor}
\draw (48.700000\du,5.400000\du)--(48.700000\du,12.800000\du)--(59.200000\du,12.800000\du)--(59.200000\du,5.400000\du)--cycle;
\pgfsetlinewidth{0.050000\du}
\pgfsetdash{}{0pt}
\pgfsetdash{}{0pt}
\pgfsetmiterjoin
\definecolor{dialinecolor}{rgb}{0.984314, 0.921569, 0.588235}
\pgfsetfillcolor{dialinecolor}
\fill (48.900000\du,6.800000\du)--(48.900000\du,7.800000\du)--(53.700000\du,7.800000\du)--(53.700000\du,6.800000\du)--cycle;
\definecolor{dialinecolor}{rgb}{0.000000, 0.000000, 0.000000}
\pgfsetstrokecolor{dialinecolor}
\draw (48.900000\du,6.800000\du)--(48.900000\du,7.800000\du)--(53.700000\du,7.800000\du)--(53.700000\du,6.800000\du)--cycle;
\pgfsetlinewidth{0.050000\du}
\pgfsetdash{}{0pt}
\pgfsetdash{}{0pt}
\pgfsetmiterjoin
\definecolor{dialinecolor}{rgb}{0.984314, 0.921569, 0.588235}
\pgfsetfillcolor{dialinecolor}
\fill (48.900000\du,8.000000\du)--(48.900000\du,9.000000\du)--(53.700000\du,9.000000\du)--(53.700000\du,8.000000\du)--cycle;
\definecolor{dialinecolor}{rgb}{0.000000, 0.000000, 0.000000}
\pgfsetstrokecolor{dialinecolor}
\draw (48.900000\du,8.000000\du)--(48.900000\du,9.000000\du)--(53.700000\du,9.000000\du)--(53.700000\du,8.000000\du)--cycle;
\definecolor{dialinecolor}{rgb}{0.000000, 0.000000, 0.000000}
\pgfsetstrokecolor{dialinecolor}
\node[anchor=west] at (48.700000\du,5.870000\du){\footnotesize{\textbf{Profile visualizations}}};
\pgfsetlinewidth{0.050000\du}
\pgfsetdash{}{0pt}
\pgfsetdash{}{0pt}
\pgfsetmiterjoin
\definecolor{dialinecolor}{rgb}{0.984314, 0.921569, 0.588235}
\pgfsetfillcolor{dialinecolor}
\fill (54.100000\du,6.800000\du)--(54.100000\du,7.800000\du)--(58.900000\du,7.800000\du)--(58.900000\du,6.800000\du)--cycle;
\definecolor{dialinecolor}{rgb}{0.000000, 0.000000, 0.000000}
\pgfsetstrokecolor{dialinecolor}
\draw (54.100000\du,6.800000\du)--(54.100000\du,7.800000\du)--(58.900000\du,7.800000\du)--(58.900000\du,6.800000\du)--cycle;
\pgfsetlinewidth{0.050000\du}
\pgfsetdash{}{0pt}
\pgfsetdash{}{0pt}
\pgfsetmiterjoin
\definecolor{dialinecolor}{rgb}{0.984314, 0.921569, 0.588235}
\pgfsetfillcolor{dialinecolor}
\fill (48.900000\du,11.600000\du)--(48.900000\du,12.600000\du)--(58.900000\du,12.600000\du)--(58.900000\du,11.600000\du)--cycle;
\definecolor{dialinecolor}{rgb}{0.000000, 0.000000, 0.000000}
\pgfsetstrokecolor{dialinecolor}
\draw (48.900000\du,11.600000\du)--(48.900000\du,12.600000\du)--(58.900000\du,12.600000\du)--(58.900000\du,11.600000\du)--cycle;
\pgfsetlinewidth{0.050000\du}
\pgfsetdash{}{0pt}
\pgfsetdash{}{0pt}
\pgfsetmiterjoin
\definecolor{dialinecolor}{rgb}{0.984314, 0.921569, 0.588235}
\pgfsetfillcolor{dialinecolor}
\fill (48.900000\du,9.200000\du)--(48.900000\du,10.200000\du)--(58.900000\du,10.200000\du)--(58.900000\du,9.200000\du)--cycle;
\definecolor{dialinecolor}{rgb}{0.000000, 0.000000, 0.000000}
\pgfsetstrokecolor{dialinecolor}
\draw (48.900000\du,9.200000\du)--(48.900000\du,10.200000\du)--(58.900000\du,10.200000\du)--(58.900000\du,9.200000\du)--cycle;
\pgfsetlinewidth{0.050000\du}
\pgfsetdash{}{0pt}
\pgfsetdash{}{0pt}
\pgfsetmiterjoin
\definecolor{dialinecolor}{rgb}{0.984314, 0.921569, 0.588235}
\pgfsetfillcolor{dialinecolor}
\fill (48.900000\du,10.400000\du)--(48.900000\du,11.400000\du)--(58.900000\du,11.400000\du)--(58.900000\du,10.400000\du)--cycle;
\definecolor{dialinecolor}{rgb}{0.000000, 0.000000, 0.000000}
\pgfsetstrokecolor{dialinecolor}
\draw (48.900000\du,10.400000\du)--(48.900000\du,11.400000\du)--(58.900000\du,11.400000\du)--(58.900000\du,10.400000\du)--cycle;
\definecolor{dialinecolor}{rgb}{0.000000, 0.000000, 0.000000}
\pgfsetstrokecolor{dialinecolor}
\node at (51.300000\du,7.300000\du){Scalasca \cite{Geimer2010}};
\pgfsetlinewidth{0.050000\du}
\pgfsetdash{}{0pt}
\pgfsetdash{}{0pt}
\pgfsetmiterjoin
\definecolor{dialinecolor}{rgb}{0.984314, 0.921569, 0.588235}
\pgfsetfillcolor{dialinecolor}
\fill (54.100000\du,8.000000\du)--(54.100000\du,9.000000\du)--(58.900000\du,9.000000\du)--(58.900000\du,8.000000\du)--cycle;
\definecolor{dialinecolor}{rgb}{0.000000, 0.000000, 0.000000}
\pgfsetstrokecolor{dialinecolor}
\draw (54.100000\du,8.000000\du)--(54.100000\du,9.000000\du)--(58.900000\du,9.000000\du)--(58.900000\du,8.000000\du)--cycle;
\definecolor{dialinecolor}{rgb}{0.000000, 0.000000, 0.000000}
\pgfsetstrokecolor{dialinecolor}
\node[anchor=west] at (49.400000\du,8.400000\du){};
\definecolor{dialinecolor}{rgb}{0.000000, 0.000000, 0.000000}
\pgfsetstrokecolor{dialinecolor}
\node at (51.300000\du,8.500000\du){CUBE \cite{Geimer2008}};
\definecolor{dialinecolor}{rgb}{0.000000, 0.000000, 0.000000}
\pgfsetstrokecolor{dialinecolor}
\node at (56.500000\du,7.300000\du){Projections \cite{Kale2006}};
\definecolor{dialinecolor}{rgb}{0.000000, 0.000000, 0.000000}
\pgfsetstrokecolor{dialinecolor}
\node at (56.500000\du,8.500000\du){Extravis \cite{Cornelissen2007}};
\definecolor{dialinecolor}{rgb}{0.000000, 0.000000, 0.000000}
\pgfsetstrokecolor{dialinecolor}
\node at (53.900000\du,9.700000\du){PerfExplorer \cite{Huck2005}};
\definecolor{dialinecolor}{rgb}{0.000000, 0.000000, 0.000000}
\pgfsetstrokecolor{dialinecolor}
\node at (53.900000\du,10.900000\du){PerformanceHat \cite{Cito2019}};
\definecolor{dialinecolor}{rgb}{0.000000, 0.000000, 0.000000}
\pgfsetstrokecolor{dialinecolor}
\node at (53.900000\du,12.100000\du){Beck et al. \cite{Beck2013}};
\pgfsetlinewidth{0.050000\du}
\pgfsetdash{}{0pt}
\pgfsetdash{}{0pt}
\pgfsetmiterjoin
\definecolor{dialinecolor}{rgb}{0.960784, 0.635294, 0.635294}
\pgfsetfillcolor{dialinecolor}
\fill (11.500000\du,11.600000\du)--(11.500000\du,12.600000\du)--(22.900000\du,12.600000\du)--(22.900000\du,11.600000\du)--cycle;
\definecolor{dialinecolor}{rgb}{0.000000, 0.000000, 0.000000}
\pgfsetstrokecolor{dialinecolor}
\draw (11.500000\du,11.600000\du)--(11.500000\du,12.600000\du)--(22.900000\du,12.600000\du)--(22.900000\du,11.600000\du)--cycle;
\definecolor{dialinecolor}{rgb}{0.000000, 0.000000, 0.000000}
\pgfsetstrokecolor{dialinecolor}
\node at (17.200000\du,12.100000\du){\textbf{This work}};
\definecolor{dialinecolor}{rgb}{0.000000, 0.000000, 0.000000}
\pgfsetstrokecolor{dialinecolor}
\node[anchor=west] at (13.400000\du,10.100000\du){};
\pgfsetlinewidth{0.050000\du}
\pgfsetdash{}{0pt}
\pgfsetdash{}{0pt}
\pgfsetmiterjoin
\definecolor{dialinecolor}{rgb}{0.984314, 0.921569, 0.588235}
\pgfsetfillcolor{dialinecolor}
\fill (21.000000\du,6.400000\du)--(21.000000\du,7.400000\du)--(26.100000\du,7.400000\du)--(26.100000\du,6.400000\du)--cycle;
\definecolor{dialinecolor}{rgb}{0.000000, 0.000000, 0.000000}
\pgfsetstrokecolor{dialinecolor}
\draw (21.000000\du,6.400000\du)--(21.000000\du,7.400000\du)--(26.100000\du,7.400000\du)--(26.100000\du,6.400000\du)--cycle;
\definecolor{dialinecolor}{rgb}{0.000000, 0.000000, 0.000000}
\pgfsetstrokecolor{dialinecolor}
\node at (23.550000\du,6.900000\du){Kerncraft \cite{Hammer2017}};
\pgfsetlinewidth{0.050000\du}
\pgfsetdash{}{0pt}
\pgfsetdash{}{0pt}
\pgfsetmiterjoin
\definecolor{dialinecolor}{rgb}{0.800000, 0.800000, 0.800000}
\pgfsetfillcolor{dialinecolor}
\fill (27.600000\du,2.800000\du)--(27.600000\du,5.000000\du)--(35.400000\du,5.000000\du)--(35.400000\du,2.800000\du)--cycle;
\definecolor{dialinecolor}{rgb}{0.000000, 0.000000, 0.000000}
\pgfsetstrokecolor{dialinecolor}
\draw (27.600000\du,2.800000\du)--(27.600000\du,5.000000\du)--(35.400000\du,5.000000\du)--(35.400000\du,2.800000\du)--cycle;
\definecolor{dialinecolor}{rgb}{0.000000, 0.000000, 0.000000}
\pgfsetstrokecolor{dialinecolor}
\node[anchor=west] at (27.600000\du,3.270000\du){\footnotesize{\textbf{Value flow analysis}}};
\pgfsetlinewidth{0.050000\du}
\pgfsetdash{}{0pt}
\pgfsetdash{}{0pt}
\pgfsetmiterjoin
\definecolor{dialinecolor}{rgb}{0.984314, 0.921569, 0.588235}
\pgfsetfillcolor{dialinecolor}
\fill (27.800000\du,3.800000\du)--(27.800000\du,4.800000\du)--(33.400000\du,4.800000\du)--(33.400000\du,3.800000\du)--cycle;
\definecolor{dialinecolor}{rgb}{0.000000, 0.000000, 0.000000}
\pgfsetstrokecolor{dialinecolor}
\draw (27.800000\du,3.800000\du)--(27.800000\du,4.800000\du)--(33.400000\du,4.800000\du)--(33.400000\du,3.800000\du)--cycle;
\definecolor{dialinecolor}{rgb}{0.000000, 0.000000, 0.000000}
\pgfsetstrokecolor{dialinecolor}
\node at (30.600000\du,4.300000\du){ValueExpert \cite{Zhou2022}};
\end{tikzpicture}